\documentclass[journal, onecolumn, 12pt]{IEEEtran}
\IEEEoverridecommandlockouts

\usepackage{cite}
\usepackage{amsmath,amssymb,amsfonts}
\usepackage{algorithmic}
\usepackage{graphicx}
\usepackage{textcomp}
\usepackage{xcolor}
\def\BibTeX{{\rm B\kern-.05em{\sc i\kern-.025em b}\kern-.08em
    T\kern-.1667em\lower.7ex\hbox{E}\kern-.125emX}}
    \usepackage{mathtools}
\usepackage{color}
\usepackage{bm}
\usepackage{amsfonts}
\usepackage{amscd}
\usepackage[mathcal]{eucal}
\usepackage{epsfig}
\usepackage{caption}
\usepackage{array}
\usepackage{eqparbox}
\usepackage{enumerate}
\usepackage{stackrel}
\usepackage{pdflscape}
\usepackage{lineno}
\usepackage{verbatim}

\usepackage{framed,xcolor,shadethm}
\usepackage{tikz}
\usetikzlibrary{calc}
    
\newcommand{\ie}{{\em i.e.}}

\newcommand{\secref}[1]{Section~\ref{#1}}
\newcommand{\figref}[1]{Fig.~\ref{#1}}

\newcommand{\thrmref}[1]{Theorem~\ref{#1}}
\newcommand{\lemref}[1]{Lemma~\ref{#1}}

\newtheorem{thrm}{\textbf{Theorem}}

\newtheorem{lemma}{\textbf{Lemma}}
\newtheorem{defn}{\textbf{Definition}}
\newtheorem{corol}{\textbf{Corollary}}

\newcommand{\abs}[1]{\left\vert#1\right\vert}
\newcommand{\norm}[1]{\Vert#1\Vert}

\def\BibTeX{{\rm B\kern-.05em{\sc i\kern-.025em b}\kern-.08em T\kern-.1667em\lower.7ex\hbox{E}\kern-.125emX}}
    
\begin{document}
\title{Secure Communication in Dynamic Wireless Ad hoc Networks}
\author{Bharath B. N. and K. G. Nagananda
\thanks{Bharath B. N. is with the Department of Electrical Engineering, Indian Institute of Technology, Dharwad, India. E-mail: \texttt{bharathbn@iitdh.ac.in}. K. G. Nagananda is with LTCI, T\'el\'ecom ParisTech, Institut Mines-T\'el\'ecom, Paris 75013, France.  E-mail: \texttt{nkyatsandra@enst.fr}. The work was supported by the Startup Grant and Networking Fund of Indian Institute of Technology, Dharwad and the European Research Council under grant agreement 715111.}
} 

\maketitle

\begin{abstract} 
We consider a wireless ad hoc network in the presence of eavesdroppers (EDs), where the nodes are distributed according to independent Poisson point processes (PPPs). The legitimate nodes follow the half-duplex mode of operation employing the slotted ALOHA protocol for transmission. For such a network, a novel communication scheme that induces a time-varying secure connectivity graph (SCG) is proposed, and the connectivity behavior of this induced SCG is studied. In particular, for a legitimate node in the network, we analyze (i) the average number of incoming edges and the average number of outgoing edges; (ii) the time to nearest-neighbor secure connectivity; and (iii) a condition on the EDs' density that allows information percolation, {\ie}, a condition for the existence of a `giant' component. The average time for secure connectivity among the nodes in this giant component is shown to scale linearly with the Euclidean distance. Further, we show that by splitting the packets into two sub-packets and routing each sub-packet along paths that are sufficiently far apart can (a) potentially improve secure connectivity and (b) reduce the overall delay incurred in exchanging packets between any two legitimate nodes in the giant component.
\end{abstract}

\section{Introduction}
Wireless ad hoc networks continue to receive significant attention due to its remarkable features of information exchange between the constituent nodes in the network. As the size of the ad hoc networks grows, the physical-layer issues especially interference prohibit the existence of a direct link between the nodes in the network. A suitable alternative is multi-hop networks, where the intermediate nodes could act as relays to improve the overall throughput. However, it has been observed that, owing to channel impairments, the time required to establish a communication link between any two nodes (path-formation time) could be longer than the number of hops in the network resulting in undesirable delays \cite{Ganti2014}. The delay analysis of wireless ad hoc networks has been of great interest in the community; see, for example, \cite{Tabet2010,Baccelli2011,Khabbaz2013,Ren2014, Ganti2014,Luo2015,Deng2017,Zhong2017,Cui2018,Li2018,Yang2018}.  

An added dimension to the delay problem arises due to eavesdroppers (EDs) with the malicious intent of trying to disrupt the communication between two nodes. When there is sufficient knowledge of the presence of EDs, the path-formation mechanism to establish connection between the source and destination nodes should also take into account the security aspect. The delay performance of wireless networks in the presence of EDs has been well reported \cite{Haenggi2008,Zhou2011,Zhou2011a,Ao2011,Pinto2012,Capar2012,Lee2012,Zhou2012,Shi2012,Gau2012,Wang2013,Zhao2013,Zhao2014a,Zhao2014,Vaze2014,Zhao2015,Zhong2018}. However, to the best of our knowledge, the characterization of average delay in secure communication using the dynamic graph model is less understood. In this paper, we make progress on this problem by studying the impact of EDs on the average delay incurred for single/multi-hop communication between a pair of nodes by considering the dynamic graph model of wireless ad hoc networks. Specifically, we consider the dynamically varying secrecy graph under a protocol similar to the one proposed in \cite{Ganti2014}. 

It is standard practice to model a dynamic wireless network as a time-varying graph; see, for example, \cite{Vaze2015}. Given a time varying wireless network, the presence and absence of links in the corresponding dynamic graph can be defined using the channel condition between any two nodes. The long-range connectivity between any two nodes in such a dynamic graph can thus be analyzed using tools from percolation theory \cite{Grimmett1999}, which is typically used to study the behavior of a connected cluster of nodes in a random graph. The existence of the so-called `giant component' is critical to get a proper handle on the connectivity of nodes in the random graph. Essentially, the giant component in a dynamic graph facilitates the study of communication between nodes located far apart, especially to characterize the time delay between the nodes involved in communication. In this paper, we concern ourselves with the analysis of time delay between nodes located in the giant component in the presence of EDs.

In our model, both legitimate nodes and EDs are assumed to be distributed according to independent Poisson Point Processes (PPPs) with different densities. The legitimate nodes follow the half-duplex mode of operation employing the slotted ALOHA protocol for transmission. We will study the impact of the transmission probability of a node on the average time required for secure-connectivity. Specifically, we will characterize the average delay of the following two packet forwarding schemes: 
\begin{enumerate}[(1)]
\item Direct-path-based approach: Two nodes can securely communicate with each other within a radius of $\eta>0$ satisfying the interference constraint with no EDs in the vicinity of the transmitter. Using slotted ALOHA this model induces a secure-connectivity graph (SCG) with time-varying links. A packet is forwarded from the source to the destination via the minimum delay path in the SCG. For a legitimate node in the network, we analyze the average number of incoming edges (in-degree) and the average number of outgoing edges (out-degree) in the static SCG, and the average time for the nearest neighbor connectivity of a node in the dynamic SCG in the presence of EDs. Different from \cite{Ganti2014}, we present a condition on the legitimate node density for the existence of a `giant component' (defined in \secref{sec:percolation}) of securely-connected links. An upper bound on the average delay incurred between any two nodes in the giant component is derived. We show that even in the presence of EDs, the average delay scales linearly with the Euclidean distance. 

\item Two secure path-based approach: Here, there exists two types of paths between two legitimate nodes in the SCG, namely, a direct path or a split path. A packet is divided into two independent sub-packets and sent to the destination along two different paths. Assume that the \emph{secure communication is achieved} if EDs cannot decode both the packets. If the two paths are ``sufficiently far apart", packets securely reach the destination. Unlike the direct-path approach, it suffices to maintain security only in the vicinity of the communicating pair. Using percolation theory, we show that the secure-connectivity of the graph can be increased, while at the same time the overall delay can be reduced for any pair of nodes in the giant component. Similar to the direct-path approach, the delay scales linearly with the Euclidean distance. This approach is similar in spirit to network coding.
\end{enumerate}

A preliminary version of this paper can be found in \cite{Bharath2019a}. Notation: $\mathbb{E}\{\cdot\}$ denotes the expectation operator, $\mathbf{1}\{\cdot\}$ is the indicator function. The Euclidian norm is denoted by $\norm{\cdot}_2$. For any $x,y \in \mathbb{R}^2$, $\mathbb{B}(x,y)$ denotes a two-dimensional Euclidian ball of radius $\norm{y}_2$ centered at $x$. We use $X \stackrel{(d)}{=} Y$ to denote that the distributions of random variables $X$ and $Y$ are same.

\secref{sec:sysmodel} comprises the system model and our protocol for secure communication. Properties of the connectivity graph resulting from the protocol model are established in \secref{sec:propgraph}. The existence of a giant component and the delay analysis for secure communication for the two schemes is presented in \secref{sec:percolation}. Concluding remarks are provided in \secref{sec:sims}.

\section{System Model} \label{sec:sysmodel}
The legitimate nodes (denoted by $\Phi_l \subseteq \mathbb{R}^2$) and the EDs (denoted by $\Phi_e\subseteq \mathbb{R}^2$) are distributed as independent PPPs with densities $\lambda_l \geq 0$ and $\lambda_e \geq 0$, respectively, in the two-dimensional Euclidean space $\mathbb{R}^2$. Time is slotted and $\Phi_l$'s employ the slotted ALOHA protocol for transmission, {\ie}, they transmit with probability $p$. Due to the half-duplex mode of operation, the set of transmitting and receiving nodes vary across time. In slot $k \in \mathbb{N}$, the set of transmitting and receiving nodes are denoted by $\Phi_t(k) \subseteq \Phi_l$ and $\Phi_r(k)\subseteq \Phi_l  \setminus \Phi_t(k)$, respectively. The following protocol decides whether two legitimate nodes can communicate with each other securely: In slot $k$, a node $x \in \Phi_t(k)$ can securely connect to a node $y \in \Phi_r(k)$ if it satisfies the following constraints:
\begin{enumerate}[(C1)]
\item for $\eta > 0$, $\norm{x-y}_2 <\eta$, which ensures that the transmitters and receivers are close enough to communicate
\item for $\beta_l > 0$, the disk $\mathbb{B}(y, \beta_l\norm{x-y}_2)$ does not contain any transmitting nodes, which ensures that a receiver does not experience interference above a tolerable level. 
\item for $\beta_e > 0$, the disk $\mathbb{B}(x,\beta_e \norm{x-y}_2)$ does not contain EDs. $\beta_e > 1$ signifies that an ED can decode the message even if it is at a farther away from the corresponding legitimate receiver. \end{enumerate}
This connectivity model together with the slotted ALOHA induces a time-varying SCG on the network. The SCG from slot $m$ to slot $n$ denoted by $G(m,n)$ is characterized by $(\Phi_l, \bigcup_{k=m}^n E_k)$, where $E_k := \{(x,y) \subseteq (\Phi_t(k),\Phi_r(k)): \mathbf{1}(x \stackrel{(1)}{\rightarrow} y) \times \mathbf{1}(\text{no EDs in }  \mathbb{B}(x,\beta_e \norm{x-y})) = 1\}$. Here $\mathbf{1}\{x \stackrel{(1)}{\rightarrow} y\}$ denotes a single hop connectivity between the two legitimate nodes $x$ and $y$ in the SCG for a given $k$ \emph{without} the security constraints. The function $\mathbf{1}(\text{no EDs in }  \mathbb{B}(x,\beta_e \norm{x-y}))= 1$ if there are no EDs in a circle of radius $\beta_e \norm{x-y}$ centered at $x$, and $0$ otherwise. A causal secure path exists between $x$ and $y$ if there exists $N>0$ time slots $k_1 < k_2 < \ldots  < k_N$ and $N$ nodes $z_1,z_2,\ldots,z_N \subseteq \Phi_l$ s.t. $(x,z_1) \in E_{k_1}$, $(z_1,z_2) \in E_{k_2},\ldots,$ $(x_N, y) \in E_{k_N}$. Two nodes are securely connected iff there is a causal path between them \cite{Ganti2014}. For SCG at a given $k$, we drop the index and let $\Phi_t$ and $\Phi_r$ denote the set of transmitting and receiving nodes, respectively.

\section{Properties of Secure-Connectivity Graph} \label{sec:propgraph}
Let us consider a legitimate node at the origin $o \equiv (0,0)$ (this is typical for PPPs \cite{Chiu2013}). For a dynamic SCG, the minimum time $T_c^S$ required for this node to securely communicate with another legitimate node in the network using a single hop is expressed as follows:
\begin{align}
T_c^S = \arg \min_k \left[\mathbf{1}(o \in \Phi_t(k))\!\!\! \prod_{x \in \Phi_r(k)}(1 - \mathbf{1}(o \stackrel{(1)}{\rightarrow} x, \mathbf{S}))\right],
\end{align}
where $\mathbf{1}(o \stackrel{(1)}{\rightarrow} x, \mathbf{S})$ denotes one hop secure connection between the node at $o \equiv (0,0)$ and the one at $x \in \mathbb{R}^2$.

\begin{lemma} \label{lem:infinite_time}
The average time for one-hop connectivity is infinity, {\ie}, $\mathbb{E}^{!o}T_C^S=\infty$. 
\label{lem:one_hop_time}
\end{lemma}
\emph{Proof}: Without the security constraint, it is known that the average time for one-hop connectivity $\mathbb{E}^{!o}[T_C^S]=\infty$, where the expectation is with respect to the palm measure \cite{Ganti2014}. Thus, with security constraint, we have $\mathbb{E}^{!o}T_C^S=\infty$. $\blacksquare$

\lemref{lem:one_hop_time} guides us in considering the average time for multi-hop connectivity. Since a multi-hop involves intermediate nodes at different time slots, it is of great interest to study the node degree distribution of the SCG at a given $k$. We next characterize the average in-degree and the out-degree of any node (for SCG at a given $k$), and the time to nearest-neighbor secure-connectivity (for dynamic SCG). Later, we will invoke these to characterize the average delay incurred in securely transmitting a packet between two legitimate nodes. 

\subsection{Average Node Degree of the SCG for a given $k$}\label{sec:nodedegreesnapshot}
Consider a transmitter at the origin, and let $N_{t,0}$ denote its secure out-degree, {\ie}, the number of nodes that it can securely connect to is given by 
\begin{eqnarray} \label{eq:Numoutdegree}
N_{t,0} := \sum_{x\in \Phi_r} \textbf{1}\{0 \stackrel{(1)}{\rightarrow} x\} \textbf{1}\{\text{no EDs in } \mathbb{B}(0, \beta_e\norm{x})\}.
\end{eqnarray}
The average out-degree is characterized by the following theorem. 

\begin{thrm} \label{thm:avg_outdegree}
For the network of legitimate nodes and EDs distributed as a  PPP with densities $\lambda_l$ and $\lambda_e$, respectively, and with legitimate nodes employing the slotted ALOHA protocol, 
the average secure out degree is given by
\begin{eqnarray} \label{thm:outdegreeStatic}
\mathbb{E} N_{t,0} = \frac{\lambda_l(1-p) [1-\exp(-(\lambda_l p \beta_l^2 + \lambda_e \beta_e^2)\pi \eta^2)]}{(\lambda_l p\beta_l^2 + \lambda_e \beta_e^2)}.
\end{eqnarray}
\end{thrm}
\emph{Proof}: See Appendix \ref{app:outdegreeProof}. $\blacksquare$

When $\lambda_e = 0$, we obtain the expression for the average out-degree without EDs derived in \cite{Ganti2014}. The average out-degree is a decreasing function of $\lambda_e$. The interference-limited regime and the noise-limited regime, which are obtained letting $\eta \rightarrow \infty$ and $\beta_l \rightarrow 0$, respectively, yield \cite{Ganti2014}
\begin{eqnarray} \label{eq:int_lim_regime_average_outdeg}
 \lim_{\eta \rightarrow \infty} \mathbb{E}N_{t,0} &=& \frac{\lambda_l (1-p)}{\lambda_l p \beta_l^2 + \lambda_e \beta_e^2,} \text{~~~~and}\\
\lim_{\beta_l \rightarrow 0} \mathbb{E}N_{t,0} &=& \frac{\lambda_l(1-p)(1-\exp(-\lambda_e \beta_e^2 \pi \eta^2))}{\lambda_e \beta_e^2}.  
\label{eq:noise_lim_regime_average_outdeg}
\end{eqnarray}

In the interference-limited regime, as $\lambda_l \rightarrow \infty$, the number of adjacent nodes and the fraction of the interfering nodes go to $\infty$. This counter-effect results in a constant average out-degree. From \eqref{eq:int_lim_regime_average_outdeg}, as $\lambda_l \rightarrow \infty$, the average out-degree $\rightarrow \frac{1-p}{p \beta_l^2}$. However, in the noise-limited regime, as $\lambda_l \rightarrow \infty$, the number of adjacent nodes $\rightarrow \infty$, and hence the average out degree also $\rightarrow \infty$ [see \eqref{eq:noise_lim_regime_average_outdeg}]. In both cases, as $\lambda_e \rightarrow \infty$, the average out-degree $\rightarrow 0$. Next, we present the average secure in-degree of a receiving node at the origin. Let $N_{r,0}$ denote the in-degree, {\ie}, the number of nodes from which the node at the origin can receive data securely:
\begin{eqnarray}
N_{r,0} := \sum_{y \in \Phi_t} \textbf{1}\{y \stackrel{(1)}{\rightarrow} 0\} \textbf{1}\{\text{no EDs in } \mathbb{B}(y, \beta_e\norm{y})\}.
 \label{eq:Numindegree}
\end{eqnarray}
The average in-degree of a typical node at the origin is given by the following theorem. 
\begin{thrm} 
For the network of legitimate nodes (employing slotted ALOHA) and EDs distributed as a  PPPs with densities $\lambda_l$ and $\lambda_e$, respectively, the average secure in-degree is given by $\mathbb{E} N_{r,0} = \frac{\lambda_lp[1-\exp(-(\lambda_l p \beta_l^2 + \lambda_e \beta_e^2)\pi \eta^2)]}{(\lambda_l p \beta_l^2 + \lambda_e \beta_e^2 )}$.
\label{thm:indegreeStatic}
\end{thrm}
\emph{Proof}: See Appendix \ref{app:indegreeProof}. $\blacksquare$\\
The expressions for the average out-degree and the average in-degree differs only by a factor of $\frac{\mathbb{E} N_{r,0}}{\mathbb{E} N_{t,0}} = \frac{p}{1-p}$. When $p=1/2$, the average in-degree is equal to the average out-degree. Thus, the observations made for the average out-degree holds good for the average in-degree as well. In the next subsection, we derive the average time required for a node in the dynamic SCG to securely connect to its nearest neighbor.

\subsection{Average Time for Secure Nearest Neighbor Connectivity}\label{sec:nearestneighborconnecttime}
Due to the slotted ALOHA, a node may require multiple attempts before it can securely connect to its neighbor. Since PPPs are assumed to be stationary, let $T_{NNC}^S$ denote the time required for a node at $(0,0)$ to securely connect to its nearest neighbor in the dynamic SCG. An expression for $\mathbb{E}\{T_{NNC}^s\}$ is given in the following theorem.
\begin{thrm}
The mean time for secure nearest neighbor connectivity in the SCG is given by
\begin{eqnarray*} 
\mathbb{E}\{T_{NNC}^S\} =  \left\{ \begin{array}{cc}
\frac{ \lambda_l}{\lambda_l - \lambda_e \beta_e^2 -\frac{p}{1-p} \lambda_l \gamma(\beta_l)}, & p<\frac{\lambda_l-\lambda_e\beta_e^2}{\lambda_l \gamma(\beta_l)- \lambda_l+ \lambda_e\beta_e^2} \\
\infty & \text{otherwise},
\label{eq:edelay}
\end{array}
\right.
\end{eqnarray*}
where 
\begin{eqnarray}
\nonumber \gamma(x) = x^2 - \frac{1}{\pi} \left\{x^2 \cos^{-1} \frac{x}{2} + \cos^{-1} \left(1 - \frac{x^2}{2}\right) - \frac{x}{2} \sqrt{4 - x^2}\right\}
\end{eqnarray}
if $x < 2$, and $x^2 - 1$ if $x \geq 2$.
\label{thm:nnsecconnect}
\end{thrm}
\emph{Proof}: See Appendix \ref{app:Proofnearestneighborseureconnectivity}. $\blacksquare$

\begin{figure}[ht]
\begin{center}
{\includegraphics[width=12.0cm]{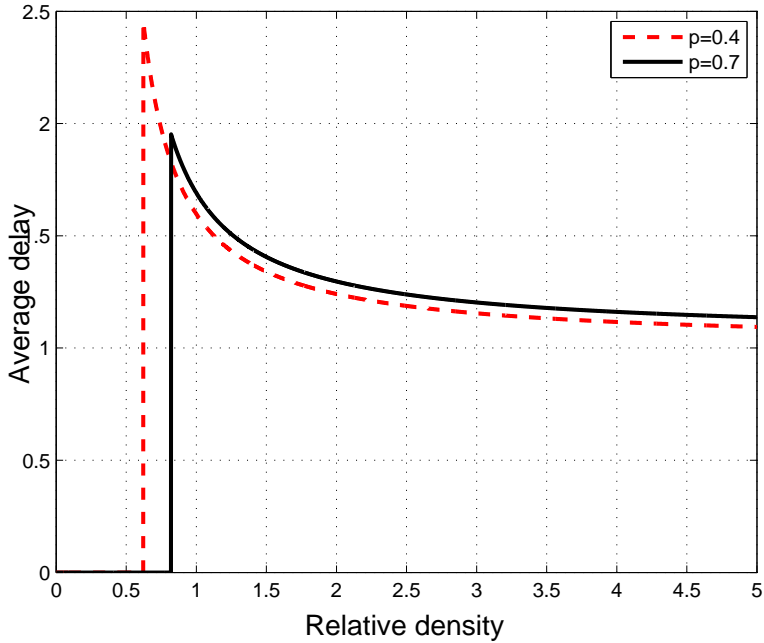}}
\caption{Average delay vs. $\lambda_l/\lambda_e$, for $\beta_e = 0.6$ and $\beta_l = 0.2$.} \label{fig:averagedelayversusRelativeDensity}
\end{center}
\end{figure}
The average time for secure nearest neighbor connectivity exhibits a phase transition behavior (see \figref{fig:averagedelayversusRelativeDensity}). As the relative density $\frac{\lambda_l}{\lambda_e}$ increases, the average delay decreases, and approaches a value surrounding $1$, which is the minimum delay possible. For higher $p$, the transition happens at higher values of the relative density, because of fewer active receivers. Next, we study the percolation and dissemination of information in the SCG for a given $k$, which will pave the way for characterizing the secure communication delay for legitimate nodes in the giant component of the dynamic SCG.

\section{Percolation of Information} \label{sec:percolation}
We begin with the definition of a giant component, {\ie}, a connected component with infinitely many nodes.
\defn{
We say that a giant component exists if \\ $\mathcal{G} = \left\{C_s \subseteq \Phi_l: \abs{C_s}=\infty,\right.$ $\left. \forall (x,y) \in C_s,~ \mathbf{1}\{x \rightarrow y,\mathbf{S}\} = 1\right\}$ occurs with probability one, where $\mathbf{1}\{x \rightarrow y,\mathbf{S}\}$ indicates the existence of a casual multi-hop secure communication between the two legitimate nodes $x$ and $y$.}\\
Even if two nodes are separated by at most $\eta$ and the interference constraint is satisfied, they may still not be able to securely communicate. This situation may arise, for example, when there are ED nodes located in the vicinity of the transmitting node, leading to the violation of condition (C3) in Sec. \ref{sec:sysmodel}. Thus, for any three consecutive nodes $x$, $y$ and $z$ in the minimum delay path, $\norm{x-z}_2$ need not be greater than $\eta$. This situation is in contrast to a wireless ad hoc network without EDs, where for the three nodes $x$, $y$ and $z$ in the minimum delay path, $\norm{x-z}_2 > \eta$; otherwise it leads to a contradiction. Thus to facilitate the proof of the existence of a giant component, and the ensuing delay analysis, we have the following definition of \emph{$\zeta$-path}, $\zeta>0$.
\defn{
A secure path is a \emph{$\zeta$-path}, if for any three consecutive legitimate nodes $x$, $y$ and $z$ in the path, {\ie}, $\norm{x-y}_2 < \eta$ and $\norm{y-z} < \eta$, we have\ $\norm{x-z}_2 > \zeta$.
}

We will now analyze the direct-path and split-path approaches for secure communication in the SCG. We derive conditions on the EDs' densities to have giant components in each of the SCGs. We prove that the paths in the giant component are $\frac{\eta}{\sqrt{3}}$-paths, which enables us to show that the delay scales linearly with the Euclidean distance between any pair of nodes in the giant component. 

\subsection{Percolation for direct-path approach} \label{subset:perc_scg_nsp}
We first derive a condition on $\lambda_l$ for a giant component to exist in the SCG. In Theorem \ref{thm:percolation}, we show that the giant component exists with non-zero probability if $\lambda_l$ scales linearly with $\lambda_e$. Once this is established, for any realization of the network, there could be potentially many paths between the nodes in the giant component. Our packet forwarding scheme picks the path with the minimum information propagation delay. The  minimum delay path need not correspond to a path with the minimum number of hops. For this protocol, we derive an expression for the average delay for the two nodes in the giant component to securely communicate (see \eqref{eq:nsp_delay}). 
\begin{thrm} 
For the \emph{secure communication protocol} model, $\exists$ $d_1 \in (0,\infty)$ and $\epsilon_{1} \in (0,1)$ s.t.
$\Pr\left\{\exists \text{ a giant component consisting of } \frac{\eta}{\sqrt{3}} \text{-paths}\right\} > 0$ if for any $0 \leq \delta < \log \left(\frac{1}{1 - \epsilon_1}\right)$, the following condition holds
\begin{align*} 
\lambda_l > \frac{\left(n_s + 2  \sqrt{3}\beta_e \right) (1+2\sqrt{3}\beta_e) \log\left[\frac{1}{1 - \exp\left\{-\delta/n_s  \right\}}\right]}{C_{\delta,\epsilon_1}} \lambda_e, 
\label{eq:percconst1}
\end{align*}
where $n_s = \left \lceil \frac{d_1 \sqrt{3}}{\eta} \right \rceil$, $C_{\delta,\epsilon_1} := \left[\log \left(\frac{1}{1 - \epsilon_1}\right) - \delta\right]$.
\label{thm:percolation}
\end{thrm}
\emph{Proof}: See Appendix \ref{app:percolation}. $\blacksquare$\\
Using \thrmref{thm:percolation} and the fact that there exists an $\eta/\sqrt{3}$-path implies the existence of a secure path. 
\begin{corol} \label{corr:nopacketsplit}
For the \emph{secure communication protocol} model, $\exists$ $d_1 \in [0, \infty]$ and $\epsilon_{1} \in [0,1]$ s.t. $\Pr\{\exists$  a giant component$\} > 0$ if the lower bound of Theorem \ref{thm:percolation} is satisfied.
\end{corol}

For $p<1$, \emph{almost surely} there exists a time slot $k<\infty$, where the interference condition is satisfied. Thus, any two legitimate nodes can communicate securely if conditions (C1) and (C3) of the communication protocol presented in Sec. \ref{sec:sysmodel} are satisfied, and if they can wait for a sufficiently long time. Due to this, the above lower bound on $\lambda_l$ does not depend on $p$ and $\beta_l$. Further, if $\lambda_e = 0$, the above bound reduces to $\lambda_l > 0$. That is, if $\lambda_l >0$, a non-zero fraction of realizations of the network will have a giant component. This is in contrast to the \emph{almost sure} existence of the giant component, which requires $\lambda_l > \frac{1.435}{\eta^2}$ \cite{Balister2010}. As $\eta \rightarrow \infty$, the lower bound on $\lambda_l$ goes to zero, which is in line with $\lambda_e=0$ case. 

We upper bound the average delay between two nodes in the giant component to analyze the variation of average time versus (i) distance, (ii) density of legitimate nodes, and (iii) density of ED nodes. The minimum delay between two nodes $x$ and $y$ is the delay between $x^* = \arg \min_{z \in \mathcal{G}} \norm{x - z}_2$ and $y^* = \arg \min_{z \in \mathcal{G}} \norm{y - z}_2$. 

Let $T^\text{(nsp)}(m,n) = \arg \min_{k \geq 0} \{k: G(0,k) \text{ has a secure causal path between }(m,0) \text{ and }(n,0)\}$ denote the minimum delay of a causal path between $(m,0)$ and $(n,0)$. $L_\text{nsp}(m,n)$ denotes the number of hops. Consider two points $(0,0)$ and $(0,n)$. The delay definition is reasonable if $\norm{x_0^* - x_n^*}_2$ scales linearly with $n$. Towards this, we have $n = \norm{(0,0) - (n,0)}_2
= \norm{(0,0) - x_0^* + x_0^* - x_n^* + x_n^* - (n,0)}_2 \stackrel{(a)}\leq \theta_1 + \norm{ x_0^* - x_n^*}_2 + \theta_2$,
where $(a)$ follows from the triangle inequality, and it follows from \cite[Lemma 8]{Kong2007} that $\theta_1 := \norm{(0,0) - x_0^*}_2$ and $\theta_2:=\norm{x_n^* - (n,0)}_2$ are \emph{almost surely} bounded random variables. Thus, $\norm{ x_0^* - x_n^*}_2 \geq n - \theta_1 - \theta_2$. Since $\theta_1$ and $\theta_2$ are bounded \emph{almost surely}, it follows from $\lim_{n \rightarrow \infty}\frac{\theta_i}{n} = 0$ \emph{almost surely} for $i=1,2$ that $\lim_{n \rightarrow \infty}\frac{\norm{ x_0^* - x_n^*}_2}{n} \geq 1$ \emph{almost surely}. 

Next, we have the following upper bound: $\norm{ x_0^* - x_n^*}_2  \leq \norm{ x_0^* - (0,0)}_2 + \norm{ (0,0) - (n,0)}_2 + \norm{(n,0)- x_n^*}_2
\leq \theta_1 + \theta_2 + n$. Similarly, we have $\lim_{n \rightarrow \infty}\frac{\norm{ x_0^* - x_n^*}_2}{n} \leq 1$ \emph{almost surely}. This implies that $\lim_{n \rightarrow \infty}\frac{\norm{ x_0^* - x_n^*}_2}{n} = 1$ \emph{almost surely}. Thus, it is sufficient to analyze the delay between two nodes along a straight line path \cite{Ganti2009}. In this analysis, we utilize the sub-additivity property of the PPP \cite{Ganti2009}
\begin{eqnarray} \label{eq:subadditive}
T^\text{(nsp)}(0,n) \leq T^\text{(nsp)}(0,k) + T^\text{(nsp)}_{T^\text{(nsp)}(0,k) }(k,n),
\end{eqnarray}
where, $T^\text{(nsp)}_{T^\text{(nsp)}(0,k) }(k,n)$ denotes the minimum delay of a causal secure path that exists between nodes at $(k,0)$ and $(n,0)$, where the causal path starts at a time slot equal to $T^\text{(nsp)}(0,k)$. Taking the expectation of \eqref{eq:subadditive} conditioned on the existence of the giant component, and using the fact that $T^\text{(nsp)}_{T^\text{(nsp)}(0,k) }(k,n) \stackrel{(d)}{=}T^\text{(nsp)}(0,n-k)$, we get $\mathbb{E}_{|\mathcal{G}}T^\text{(nsp)}(0,n) \leq \mathbb{E}_{|\mathcal{G}}T^\text{(nsp)}(0,k) + \mathbb{E}_{|\mathcal{G}}T^\text{(nsp)}(0,n-k)$. Recursively applying this, we get $\mathbb{E}_{|\mathcal{G}} T^\text{(nsp)}(0,n) \leq n \mathbb{E}_{|\mathcal{G}} T^\text{(nsp)}(0,1)$. Thus, to upper bound the average delay, it suffices to upper  bound $\mathbb{E}_{|\mathcal{G}}  T^\text{(nsp)}(0,1)$. To do this, we simply upper bound the average number of hops between nodes at $(0,0)$ and $(1,0)$:
\begin{thrm} 
For our protocol model, and given the conditions of Theorem \ref{thm:percolation}, $\mathbb{E}_{|\mathcal{G}} L_\text{nsp}(0,1)$ can be upper-bounded as 
\begin{eqnarray} \label{eq:delay1}
\mathbb{E}_{|\mathcal{G}} L_\text{nsp}(0,1) \leq \frac{12 d^2_{\delta_{\text{nsp}}}}{\pi \eta^2} \left( \frac{ 9 {\delta^4_{\text{nsp}}}}{9 {\delta^4_{\text{nsp}}} - 8}\right) + \frac{1 + \delta_{\text{nsp}}^4}{\delta_{\text{nsp}}^4}
\end{eqnarray}
for some $\delta_{\text{nsp}} > \sqrt[\leftroot{-2}\uproot{2}4]{\frac{8}{9}}$ and $d_{\delta_{\text{nsp}}} < \infty$. Also, $\mathbb{E}_{|\mathcal{G}} L_\text{nsp}(0,1) < \infty$. 
\label{thm:meanhopnosplit}
\end{thrm}
\emph{Proof}: See Appendix \ref{app:meanhopnosplit}. $\blacksquare$\\
\thrmref{thm:meanhopnosplit} provides an upper bound on the average number of hops between a pair of nodes in the giant component that are close to the origin and $(0,1)$. An upper bound on the corresponding average delay is at most equal to the sum of the delays of each of the links in the path. A more precise statement is proved in the following lemma.

\begin{lemma} 
For our protocol model, given the conditions of Theorem \ref{thm:percolation}, for some $\delta_{\text{nsp}} > \sqrt[\leftroot{-2}\uproot{2}4]{\frac{8}{9}}$ and $d^2_{\delta_{\text{nsp}}} \geq d$, we have
\begin{eqnarray}
\nonumber \mathbb{E}_{|\mathcal{G}} T_\text{nsp}(0,1) \leq \exp\left(\frac{\eta^2}{3}\left(\lambda_e \pi \beta_e^2  + \lambda_l \frac{p}{1-p} \pi \gamma(\beta_l)\right)\right) \left[\frac{12 d^2_{\delta_{\text{nsp}}}}{\pi \eta^2} \left( \frac{ 9 {\delta^4_{\text{nsp}}}}{9 {\delta^4_{\text{nsp}}} - 8}\right) + \frac{1 + \delta_{\text{nsp}}^4}{\delta_{\text{nsp}}^4}\right].
 \label{eq:average_delay1}
\end{eqnarray}
\label{thm:append_average_delay1}
\end{lemma}
\emph{Proof:} See Appendix \ref{sec:append_average_delay1}. $\blacksquare$

\lemref{thm:append_average_delay1} and $\mathbb{E}_{|\mathcal{G}} T_\text{nsp}(0,n) \leq n \mathbb{E} T_\text{nsp}(0,1)$ yields
\begin{eqnarray*}
\nonumber \frac{\mathbb{E}_{|\mathcal{G}} T_\text{nsp}(0,n)}{n}   \leq \exp\left(\frac{\eta^2}{3}\left(\lambda_e \pi \beta_e^2  + \lambda_l \frac{p}{1-p} \pi \gamma(\beta_l)\right)\right) \left[\frac{12 d^2_{\delta_{\text{nsp}}}}{\pi \eta^2} \left( \frac{ 9 {\delta^4_{\text{nsp}}}}{9 {\delta^4_{\text{nsp}}} - 8}\right) + \frac{1 + \delta_{\text{nsp}}^4}{\delta_{\text{nsp}}^4}\right],
 \label{eq:nsp_delay}
\end{eqnarray*}
where the right hand side is finite for $p < 1$. Therefore, for $\lambda_e < \infty$, an upper bound on the average delay between any pair of nodes in the giant component scales linearly with the Euclidean distance. 

\begin{lemma} \label{lem:delay_opt}
A lower bound on the average delay between two nodes that are separated by a distance $d$ scales linearly with $d$, i.e., 
$\mathbb{E}_{|\mathcal{G}} T^{\text{opt}}(0,d) \geq \left(\frac{d}{\eta} - 1\right) \exp\left({\eta^2}\left( \lambda_l \frac{p}{1-p} \pi \gamma(\beta_l)\right)\right)$, 
where $T^{\text{opt}}(0,d)$ is the delay for the optimum scheme. 
\end{lemma}
Thus, $ \lim_{d \rightarrow \infty} \frac{\mathbb{E}T^\text{nsp}(0,d)}{d} = \text{constant} < \infty$, concluding that the scaling of the average delay with the Euclidian distance between source and destination nodes in the giant component remains unchanged due to the presence of EDs. 

\begin{figure}[ht]
  \begin{center}
  {\includegraphics[height = 10cm, width=12cm]{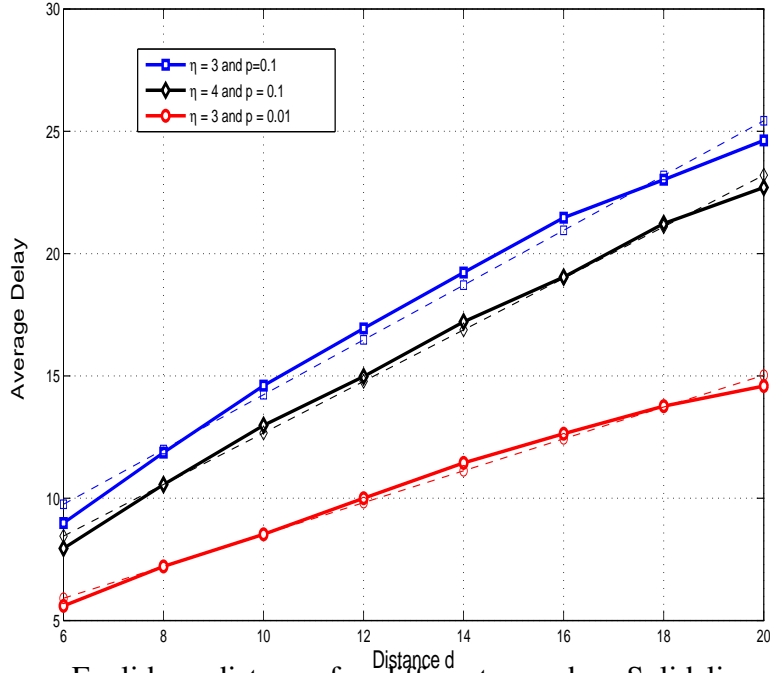}}
 \vspace{-0.8cm} \caption{Average delay vs. Euclidean distance for different $p$ and $\eta$. Solid lines refer to the simulation, dashed line refers to a linear fit. $\beta_e = 0.8$, $\beta_l = 1.2$.} 
 \label{fig:delay_vs_dist}
  \end{center}
  \end{figure}

In \figref{fig:delay_vs_dist}, we show the average delay for the direct path approach versus distance for different values of $p$ and $\eta$. The legitimate nodes and EDs are located in a $20$m$^2$ area distributed according to PPPs with $\lambda_l = 1$ and $\lambda_e = 0.1$, respectively. We have also plotted the straight line approximation along with the actual delay. The delay is averaged over $2000$ realizations of the network. As seen, the delay scales linearly with the Euclidean distance.
Further, the right hand side of \eqref{eq:nsp_delay} is a monotonically increasing function of $p$. This is because as $p$ increases, more legitimate nodes become transmitters leading to a larger time for per hop communication. Similar conclusion is drawn when $\lambda_e = 0$ \cite{Ganti2009}. As $\eta$ increases, the communication range of each legitimate node increases. This leads to longer hops, and hence requires stringent condition on the interference [see (C2) of Sec. \ref{sec:sysmodel}], leading to a larger per hop delay. On the other hand, increasing $\eta$ decreases the number of hops between any pair of legitimate nodes in the giant component. This tradeoff is captured in \eqref{eq:nsp_delay}. The direct-path protocol has the limitation that a shorter path may not always exist due to the presence of EDs leading to a larger delay. Next, we develop the split-path approach based on the idea of packet splitting \cite{Capar2012} to improve the connectivity and the delay performance compared to the direct-path scheme.

\subsection{Percolation for split-path approach} \label{sec:packetsplit}
One of the drawbacks of the direct-path scheme is that the entire path in the SCG needs to be secure. This leads to a stringent requirement on the number of legitimate nodes that enable percolation of information. To alleviate this shortcoming, we make the following assumption.\\
\textbf{Assumption 1:} Every packet in the network can be broken into two independent packets. Secure communication takes place if none of the EDs is able to decode both the packets. 

For any pair of legitimate nodes in the network, packets are routed along a direct-path, or a spit-path, or a combination of direct and split paths. The main idea behind split-path (packet-splitting) algorithm is to break the packet into two sub-packets. By \textbf{Assumption 1}, no ED should be able to decode both the packets. This amounts to routing the two packets in two different paths that are ``sufficiently" far apart to the destination node maintaining the security constraint only at a region ``near" the source node and the destination node. Thus, the entire region in the path need not be secure. It suffices for the source-destination pair to know the locations of EDs in its neighborhood. We use the term ``two-secure path" for the path which consists of a combination of both direct and split-paths. We first establish a condition for the existence of a giant component corresponding to the packet splitting scheme. 
\begin{thrm} \label{thm:percolation_splitting}
For our protocol model with splitting and routing of packets, 
there exists $d_2 \in (0, \infty)$ and $\epsilon_{2} \in (0,1)$ such that $\Pr\{\exists \text{ a giant component}\} > 0$, 
if for any $0 \leq \delta < \log \left(\frac{1}{1 - \epsilon_2}\right)$,
$\lambda_l \geq \min\{\mathcal{\rho}_\text{sp},\mathcal{\rho}_\text{nsp}\} \lambda_e$, 
where 
$\mathcal{\rho}_\text{sp}:= \frac{4 \left(1+ 2 \sqrt{3} \beta_e\right) (4 +  \sqrt{3} \beta_e) }{C_{\delta, \epsilon_2}} \log\left(\frac{1}{1 - \exp\left\{-\frac{\delta}{2 \left(n_s  + 4 \right)}\right\} } \right)$
and
$\mathcal{\rho}_\text{nsp} : = \frac{\left(n_s + 2  \sqrt{3}\beta_e \right) (1+2\sqrt{3}\beta_e) }{C_{\delta,\epsilon_2}} \log\left(\frac{1}{1 - \exp\left\{-\frac{\delta}{n_s}\right\} } \right).$
Here, $n_s:= \left \lceil \frac{d_2 \sqrt{3}}{\eta} \right \rceil$ and $C_{\delta,\epsilon_2} := \left[\log \left(\frac{1}{1 - \epsilon_2}\right) - \delta\right]$.
\end{thrm}
\emph{Proof}: See Appendix \ref{app:percolation_splitting} for the proof. $\blacksquare$


As before, let the minimum delay between any two points $x$ and $y$ in $\mathbb{R}^2$ be the delay between the nodes $x^*$ and $y^*$ in the giant component that are close to $x$ and $y$, respectively. We let $T^\text{(sp)}(m,n)$ to denote the minimum delay of a causal path between $(m,0)$ and $(n,0)$, and the corresponding number of hops by $L_\text{sp}(m,n)$. With a slight abuse of notation, let $\mathcal{G}$ denote the giant component. Also, $\mathbb{E}_{|\mathcal{G}} T^\text{(sp)}(0,n) \leq n \mathbb{E}_{|\mathcal{G}}T^\text{(sp)}(0,1)$. Thus, to obtain an upper bound on the average delay $T^\text{(sp)}(0,n)$, it suffices to upper  bound $\mathbb{E}_{|\mathcal{G}}  T^\text{(sp)}(0,1)$. In fact, it suffices to upper bound $\mathbb{E}_{|\mathcal{G}} L_\text{sp}(0,1)$, as shown below.
\begin{thrm} 
For our protocol model with the splitting and routing of packets, and given the conditions of Theorem \ref{thm:percolation_splitting}, we have $\mathbb{E}_{|\mathcal{G}} L_\text{sp}(0,1) \leq \frac{12 d^2_{\delta_{\text{sp}}}}{\pi \eta^2} \left( \frac{9 {\delta^4_{\text{sp}}}}{9 {\delta^4_{\text{sp}}} - 8}\right) + \frac{1 + \delta_{\text{sp}}^4}{\delta_{\text{sp}}^4}$, for some $\delta_{\text{sp}} > \sqrt[\leftroot{-2}\uproot{2}4]{\frac{8}{9}}$ and $d_{\delta_{\text{sp}}} < \infty$. Further, $\mathbb{E}_{|\mathcal{G}} L_\text{sp}(0,1) < \infty$. 
\label{thm:meanhopsplit}
\end{thrm}
\emph{Proof}: See Appendix \ref{app:meanhopsplit}. $\blacksquare$

Lastly, we derive an upper bound on the average delay between any two nodes in the giant component. 
\begin{corol} \label{thm:average_delay2}
For the protocol model in Sec. \ref{sec:sysmodel}, and given the conditions of Theorem \ref{thm:percolation_splitting}, an upper bound on the average delay denoted $\mathbb{E}_{|\mathcal{G}} T_\text{sp}(0,1)$ between any two nodes in 
the giant component is given by
\begin{eqnarray} 
\nonumber \mathbb{E}_{|\mathcal{G}} T_{\text{sp}}(0,1) \leq \exp\left[{\eta^2}\left(\lambda_e \pi \beta_e^2  + \lambda_l \frac{p}{1-p} \pi \gamma(\beta_l)\right)\right] \left[\frac{12 d^2_{\delta_{\text{sp}}}}{\pi \eta^2} \left( \frac{ 9 {\delta^4_{\text{sp}}}}{9 {\delta^4_{\text{sp}}} - 8}\right) + \frac{1 + \delta_{\text{sp}}^4}{\delta_{\text{sp}}^4}\right]
\label{eq:average_delay2}
\end{eqnarray}
for some $\delta_{\text{sp}} > \sqrt[\leftroot{-2}\uproot{2}4]{\frac{8}{9}}$ and $d_{\delta_{\text{sp}}} < \infty$. Further, $\mathbb{E}_{|\mathcal{G}} T_{\text{sp}}(0,1) < \infty$.  
\end{corol}
\emph{Proof:} Similar to that of Theorem  \ref{thm:append_average_delay1}, and is omitted. $\blacksquare$ \\
A combination of direct and split paths is more resilient to EDs since $\lambda_e$ has lesser impact on the delays corresponding to split-paths. The above bound indirectly captures this effect through $\delta_\text{sp}$, which is the crossing probability of a rectangle of size $d_{\delta_\text{sp}}$. Intuitively, the rate at which $\delta_\text{sp}$ decreases with $\lambda_e$ is faster compared to the rate at which $\delta_\text{nsp}$ decreases with $\lambda_e$. Now, $\mathbb{E}_{|\mathcal{G}} T_\text{sp}(0,n) \leq n \mathbb{E}_{|\mathcal{G}} T_\text{sp}(0,1)$ implies that 
\begin{eqnarray}
\nonumber \frac{\mathbb{E}_{|\mathcal{G}} T_\text{sp}(0,n)}{n} \leq  \exp\left[{\eta^2}\left(\lambda_e \pi \beta_e^2  + \lambda_l \frac{p}{1-p} \pi \gamma(\beta_l)\right)\right]  \left[\frac{12 d^2_{\delta_{\text{sp}}}}{\pi \eta^2} \left( \frac{ 9 {\delta^4_{\text{sp}}}}{9 {\delta^4_{\text{sp}}} - 8}\right) + \frac{1 + \delta_{\text{sp}}^4}{\delta_{\text{sp}}^4}\right],
\end{eqnarray}
where the right hand side is independent of $n$, and hence finite. This along with the lower bound in Lemma \ref{lem:delay_opt} shows that the average delay scales linearly with the Euclidian distance between any two legitimate nodes in the giant component. 

This result utilizes the fact that the per hop delay is at most equal to $\exp\left({\eta^2}\left(\lambda_e \pi \beta_e^2  + \lambda_l  \frac{p}{1-p} \pi \gamma(\beta_l)\right)\right)$. The bound does not capture the per-hop delays involved in the split-paths. Accounting for the average delay corresponding to split-paths amounts to finding the average number of hops in the split-path along any two-secure paths, which is mathematically intractable. However, this bound captures the scaling behavior of the delay with respect to the Euclidian distance. The differences lie in the condition for the giant component to exist and the constants in the delay expressions. 

To contrast the two schemes, we compare the results of Theorems \ref{thm:percolation} and \ref{thm:percolation_splitting} with those of Theorems \ref{thm:meanhopsplit} and \ref{thm:meanhopnosplit}, respectively. Firstly, $\mathcal{S_\text{nsp}} \subseteq \mathcal{S_\text{sp}}$, where $\mathcal{S_\text{nsp}}$ and $\mathcal{S_\text{sp}}$ denote the existence of a giant component for direct-path scheme and two-secure path scheme, respectively.
\begin{lemma} \label{lem:discussion_compare}
Given the schemes, $\Pr\{\mathcal{S_\text{nsp}}\} \leq \Pr\{\mathcal{S_\text{sp}}\}$ and $\mathbb{E}_{|\mathcal{G}} T_\text{sp}(0,n) \leq \mathbb{E}_{|\mathcal{G}} T_\text{nsp}(0,n)$.
\end{lemma}
Since $\mathcal{S_\text{nsp}} \subseteq \mathcal{S_\text{sp}}$, the minimum $\lambda_l$ required for the giant component to exist is smaller with packet splitting than with direct-path. Therefore, from the connectivity viewpoint, packet splitting is better than direct-path, {\ie}, given $\lambda_l$ and $\lambda_e$, the number of secure connections is larger on an average for packet splitting. To compare the average delay for secure communication, we let the crossing probabilities of the two schemes to be equal, {\ie}, $\delta_\text{sp} = \delta_\text{nsp}$. Thus, to show the significance of the packet splitting approach, it suffices to prove that $d_{\delta_\text{sp}} \leq d_{\delta_\text{nsp}}$. The proof follows from the fact that $\mathcal{V_\text{nsp}} \subseteq \mathcal{V_\text{sp}}$, where $\mathcal{V_\text{sp}}$ and $\mathcal{V_\text{nsp}}$ denote the number of securely connected links in the network with and without packet splitting, repspectively. We conclude that the splitting-based approach is better compared to direct-path. Both the schemes are order-optimal with respect to the Euclidian distance, {\ie}, the delays in both cases scale linearly with Euclidean distance. The performance gain obtained by using packet splitting approach is only in terms of a constant factor.

\section{Concluding remarks}\label{sec:sims}
This paper considered the delay analysis of a wireless ad hoc network in the presence of EDs. We assumed that both legitimate nodes and ED nodes are distributed according to independent PPPs. The legitimate nodes employ slotted ALOHA for packet transmission. The secure communication protocol is shown to induce a dynamic SCG. We analyzed certain characteristics of this static/dynamic SCG such as the average in-degree and out-degree, and the average time for the nearest neighbor secure-connectivity. For the SCG, we derived a condition for the existence of a giant component. An upper bound on the average delay for the two nodes in the giant component to securely exchange information was derived. The upper bound reveals that the average delay scales linearly with the Euclidean distance between any pair of nodes in the giant component. Further, we showed that by breaking the packets into two sub-packets, and routing the sub-packets packets in different paths that are sufficiently far apart can potentially improve the secure-connectivity and reduce the overall delay incurred in exchanging packets between any two nodes in the giant component. For this approach, the delay again scales linearly with the Euclidian distance. Thus, we both the proposed approaches are order optimal with respect to the Euclidian distance, while the packet-splitting approach can improve the delay by only a constant scaling factor. The study of the impact of node mobility and directional antennas for secure communication are interesting avenues for future research.


\appendix

\subsection{Proof of \thrmref{thm:avg_outdegree}}  \label{app:outdegreeProof}
From \eqref{eq:Numoutdegree}, we have $\mathbb{E} N_{t,0} = \mathbb{E}_{\Phi_l} \mathbb{E}_{\Phi_e} \sum_{x\in\Phi_r} \textbf{1}\{0 \stackrel{(1)}{\rightarrow} x\} \times \textbf{1}\{\text{no EDs in }\ \mathbb{B}(0, \beta_e\norm{x})\}$. Using the fact that 
$\mathbb{E}\textbf{1}\{\text{no EDs in }\ \mathbb{B}(0, \beta_e\norm{x})\} = \Pr\left\{\text{no EDs in }\right.$ $\left.\mathbb{B}(0, \beta_e\norm{x})\right\},$
which is equal to first contact distribution of the PPP, the expression simplifies to $\mathbb{E}_{\Phi_l} \sum_{x\in \Phi_r} \textbf{1}\{0 \stackrel{(1)}{\rightarrow} x\} \exp (-\lambda_e \pi \beta_e^2 \norm{x}^2)$. Using Campbell's theorem \cite{Chiu2013}, we get
\begin{eqnarray}
\lambda_l(1-p) \int_{\mathbb{B} (0,\eta)} \exp(-\lambda_l \beta_l^2 \pi \norm{x}^2) \exp(-\lambda_e \pi \beta_e^2 \norm{x}^2) dx 
&=& \frac{\lambda_l(1-p) E_g}{(\lambda_l p\beta_l^2 + \lambda_e \beta_e^2)},
\end{eqnarray}
where $E_g := [1-\exp(-(\lambda_l p \beta_l^2 + \lambda_e \beta_e^2)\pi \eta^2)]$. This completes the proof of \thrmref{thm:avg_outdegree}.
$\blacksquare$

\subsection{Proof of \thrmref{thm:indegreeStatic}}  \label{app:indegreeProof}
Taking the expectation of \eqref{eq:Numindegree}, $\mathbb{E}N_{r,0}$ can be written as 
\begin{eqnarray}
 \mathbb{E}N_{r,0} &=& \mathbb{E}_{\Phi_l,\Phi_e}\sum_{y\in \Phi_t}  \textbf{1}\{ y  \stackrel{(1)}{\rightarrow} 0\} \times \textbf{1}\{\text{no EDs in } \mathbb{B}(y,\beta_e\norm{y})\} \nonumber \\
                  &\stackrel{(a)}{=}& \mathbb{E}_{\Phi_l} \sum_{y\in\Phi_t} \textbf{1}\{y  \stackrel{(1)}{\rightarrow}  0\} \exp(-\lambda_e \pi \beta_e^2 \norm{y}^2)\nonumber \\
                  &\stackrel{(b)}{=}& \lambda_l p \int_{\mathbb{R}^2} \mathbb{E}_{\Phi_t}[\textbf{1}\{y  \stackrel{(1)}{\rightarrow} 0\}]\exp(-\lambda_e \pi \beta_e^2 \norm{y}^2) dy, \nonumber \\
&=& \lambda_l p \int_{\mathbb{B}(0,\eta)} \exp(-\lambda_l p \pi \beta_l^2 \norm{y}^2 - \pi \lambda_e \beta_e^2 \norm{y}^2 )dy,  \label{eq:thm2}
\end{eqnarray}
where (a) follows from the first contact distribution of the PPP, and (b) follows from the Campbell's theorem \cite{Chiu2013}. Solving for equation \eqref{eq:thm2} proves \thrmref{thm:indegreeStatic}. $\blacksquare$

\subsection{Proof of \thrmref{thm:nnsecconnect}}  \label{app:Proofnearestneighborseureconnectivity}
Let $N_{g,0} = z$ be the nearest neighbor to the origin. Let us denote the indicator function corresponding to a one-hop secure connection between the node at the origin and the nearest neighbor at $z$ after $k$ time slots by
\begin{eqnarray}
\textbf{1}_k\{0  \stackrel{(1)}{\rightarrow} z, \mathbf{S}\} = \prod_{x \in \mathcal{F}} \left(1-\textbf{1}\{x\in \mathbb{B}(z,\beta_l\norm{z})\} \textbf{1}\{x\in \Phi_t(k)\}\right)\times \textbf{1} \{\text{no EDs in } \mathbb{B}(0,\beta_l \norm{z})\},
\end{eqnarray}
where $\mathcal{F} := \Phi_l \cap \mathbb{B}(0, \norm{z})^c$.
Since $T_{NNC}^S$ is a positive random variable, we can write
\begin{eqnarray}
\mathbb{E}\{T_{NNC}^S\} = \sum_{k=0}^\infty \mathbb{P}\{T_{NNC}^S > k\} 
= \mathbb{E}_{z,\Phi_l} \sum_{k=0}^\infty \mathbb{P}\{T_{NNC}^S > k | z, \Phi_l\}. \label{eq:averageNNC2}
\end{eqnarray}
But, $\mathbb{P}\{T_{NNC}^S > k | z, \Phi_l\} = \mathbb{E}_{\Phi_e} \prod_{l=0}^k (1-\textbf{1}_l\{0  \stackrel{(1)}{\rightarrow} z, \mathbf{S}\})$, which can be used in \eqref{eq:averageNNC2} to get
\begin{eqnarray}
\mathbb{E}\{T_{NNC}^S\} &=& \mathbb{E}_{Z,\Phi_l} \sum_{k=0}^\infty \mathbb{E}_{\Phi_e} \prod_{l=0}^k (1-\textbf{1}_l\{0  \stackrel{(1)}{\rightarrow} z, \mathbf{S}\}) \nonumber \\
&\stackrel{(a)}{=}& \mathbb{E}_{Z,\Phi_l} \sum_{k=0}^\infty s_k \nonumber 
= \mathbb{E}_Z {\exp(\lambda_e \pi \beta_e^2 \norm{z}^2)}\mathbb{E} \prod_{x\in \Phi_l \cap \mathbb{B}(0,\norm{z})^c} f(x,z) \nonumber \\
&\stackrel{(b)}{=}& \mathbb{E}_Z \exp\left(\lambda_e \pi \beta_e^2 \norm{z}^2\right) \exp\left(-\lambda_l g(z)\right) \nonumber \\
\nonumber &=& \mathbb{E}_Z \exp\left(\lambda_e \pi \beta_e^2 \norm{z}^2 + \lambda_l \frac{p}{1-p} \pi \gamma(\beta_l)\norm{z}^2\right) \\
&=& 2\pi \lambda_l \int_0^\infty y \exp\left[-\left(\lambda_l - \lambda_e\beta_e^2 - \frac{p}{1-p} \lambda_l \gamma(\beta_l) \right)y^2\right] dy, \label{eq:averagedelayzdistance}
\end{eqnarray}
where (a) follows from the fact that the ALOHA protocol is independent across time,
\begin{eqnarray}
s_k &=& \left[1- \prod_{x\in \Phi_l \cap \mathbb{B}(0,\norm{z})^c} (1-\textbf{1}\{x \in \mathbb{B}(z,\beta_l\norm{z})\}p)\exp(-\lambda_e \pi \beta_e^2 \norm{z}^2)\right]^k,\\
f(x,z) &=& \frac{1}{(1-\textbf{1}\{ x \in \mathbb{B}(z,\beta_l\norm{z})\}p)}, \\ 
g(z) &=& \int_{\mathbb{B}(0, \norm{z})^c} \left[1-\frac{1}{1-\textbf{1}\{x\in \mathbb{B}(z,\beta_l  \norm{z})\}p}\right] dx, 
\end{eqnarray}
and 
(b) follows from the probability generating functional \cite{Chiu2013}, and $\gamma(\beta_l)$ is defined in \thrmref{thm:nnsecconnect}. Finally, \eqref{eq:averagedelayzdistance} can be simplified to get \eqref{eq:edelay} which completes the proof of \thrmref{thm:nnsecconnect}.
$\blacksquare$

\subsection{Proof of \thrmref{thm:percolation}} \label{app:percolation}
Recall that $\{\mathbf{1}_\omega (x\rightarrow y,\mathbf{S})\}$ denotes the existence of a secure $\omega$-path between the two legitimate nodes $x$ and $y$. For the sake of contradiction, suppose that there is no percolation \cite{Vaze2015}, {\ie},
\begin{eqnarray} \label{eq:percapp}
\Pr\{\exists ~ C \subseteq \Phi_l : \abs{C} = \infty~\text{and}~\forall(x,y)\subseteq C,~~\mathbf{1}_s (x\rightarrow y,\mathbf{S})=1\} = 0,
\end{eqnarray}
where $s>0$ (to be determined later). Later we will choose $s = \frac{\eta}{\sqrt{3}}$ as required. Here, a path always refers to a secure $s$-path, unless otherwise stated explicitly.
Denote a ball centered at the origin $o \equiv (0,0)$ and radius $n\in \mathbb{N}$ by $\mathbb{B}(0,n)$. We say that the ball $\mathbb{B}(0,n)$ is open if $\forall$ $x \in \mathbb{B}(0,n) \cap \Phi_l$, there exists at least one node $y \in \mathbb{B}(0,n)^c \cap \Phi_l$  such that $\mathbf{1}_s\{x\rightarrow y, \mathbf{S}\} = 1$; otherwise, the ball is closed. From \eqref{eq:percapp}, there exists an $i \in \mathbb{N}$ such that $\mathbb{B}(0,i)$ is closed. Therefore, we have 
\begin{eqnarray}
\nonumber 1 &=& \Pr\{\cup_{i=1}^\infty \mathbb{B}(0,i) \text{ is closed}\} \\
\nonumber &=& \Pr\{\lim_{n \rightarrow \infty} \bigcup_{i=1}^n \mathbb{B}(0,i) \text{ is closed}\}\\
&=& \lim_{n \rightarrow \infty} \Pr\{ \bigcup_{i=1}^n \mathbb{B}(0,i) \text{ is closed}\}, 
\end{eqnarray}
where the last equality follows from the continuity property of the probability measure. Using the limit property, this implies that there exists an $\bar \epsilon \in (0,1)$ and $N_{\bar \epsilon} < \infty$ such that $\Pr\{ \bigcup_{i=1}^{N_{\bar \epsilon}} \mathbb{B}(0,i) \text{ is closed}\} > \bar \epsilon$. 

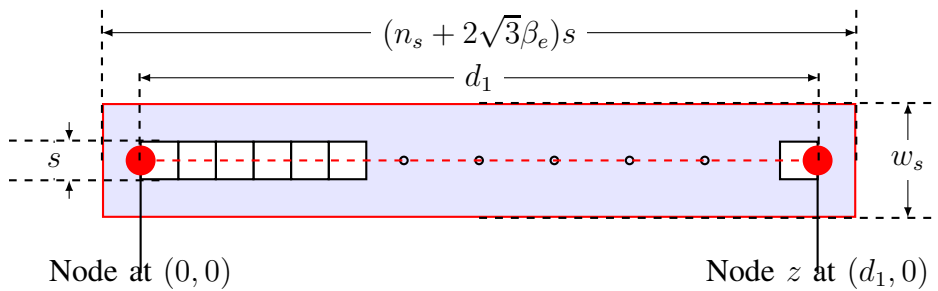
\begin{figure}[ht] 
  \begin{center}
\begin{tikzpicture}[thick]
\tikzset{%
    dimen/.style={<->,>=latex,thin,every rectangle node/.style={fill=white,midway,font=\sffamily}},
    symmetry/.style={dashed,thin},
}
\tikzstyle{lightGrayBoxStyle}=[fill=white!20,draw=black,minimum height=0.5cm,minimum width=0.5cm]
\tikzstyle{outerbox}=[fill= white!90!blue,draw=red,minimum height=1.5cm,minimum width=10cm]
\node[outerbox] at (5,0.75) (o1) {};
\node[lightGrayBoxStyle] at (9.25,0.75) (b7) {};
\node[lightGrayBoxStyle] at (3.25,0.75) (b6) {};
\node[lightGrayBoxStyle] at (2.75,0.75) (b5) {};
\node[lightGrayBoxStyle] at (2.25,0.75) (b4) {};
\node[lightGrayBoxStyle] at (1.75,0.75) (b3) {};
\node[lightGrayBoxStyle] at (1.25,0.75) (b2) {};
\node[lightGrayBoxStyle] at (0.75,0.75) (b1) {};
\draw (4,0.75) circle (0.05cm);
\draw (5,0.75) circle (0.05cm);
\draw (6,0.75) circle (0.05cm);
\draw (7,0.75) circle (0.05cm);
\draw (8,0.75) circle (0.05cm);
\draw[red,thick,dashed] (0.5,0.75) -- (9.5,0.75);
\draw[black,thick] (0.5,0.75) -- (0.5,-0.75) node {Node at $(0,0)$};
\draw[black,thick] (9.5,0.75) -- (9.5,-0.75) node {Node $z$ at $(d_1,0)$};
\draw (0.5,0.75) node[circle,fill=red]{} circle (0.01cm);
\draw (9.5,0.75) node[circle,fill=red]{} circle (0.01cm);
\draw [dimen,-] ([xshift=-1.4cm]b1.south) -- ([xshift=-1.4cm]b1.north) node  {$s$};
\draw [dimen,<-,rotate=-90] ([yshift=-1.2cm]b1.north) -- ++(-7pt,0);
\draw [dimen,<-,rotate=90] ([yshift=1.2cm]b1.south) -- ++(-7pt,0);
\draw[dashed] ([xshift=-0.0cm]b1.south) --([xshift=-2.0cm]b1.south);
\draw[dashed] ([xshift=-0.0cm]b1.north) --([xshift=-2.0cm]b1.north);

\draw [dimen,<->] ([xshift=5.7cm]o1.south) -- ([xshift=5.7cm]o1.north) node  {$w_s$};
\draw [dimen,<->] ([yshift=1.1cm]b1.west) -- ([yshift=1.1cm] b7.east) node {$d_1$};
\draw [dimen,<->] ([yshift=1.7cm]o1.west) -- ([yshift=1.7cm] o1.east) node {$(n_s + 2\sqrt{3} \beta_e)s$};
\draw[dashed] ([yshift=0.0cm]b1.west) --([yshift=1.25cm]b1.west);
\draw[dashed] ([yshift=0.0cm]b7.east) --([yshift=1.25cm]b7.east);
\draw[dashed] ([yshift=0.0cm]o1.east) --([yshift=2.10cm]o1.east);
\draw[dashed] ([yshift=0.0cm]o1.west) --([yshift=2.10cm]o1.west);
\draw[dashed] ([xshift=0.0cm]o1.north) --([xshift=6.30cm]o1.north);
\draw[dashed] ([xshift=0.0cm]o1.south) --([xshift=6.30cm]o1.south);
\end{tikzpicture}
\caption{Nodes at o can securely communicate if the region bounded by the rectangle of dimension $w_s \times (n_s + 2\sqrt{3} \beta_e) s$ does not 
have any EDs and each tiles of size $s\times s$ at the center contains at least one legitimate node. Here, $w_s:= (1+ 2\sqrt{3} \beta_e)s$.}\label{fig:percnosplit}
  \end{center}
\end{figure}

From the union bound, we have $\sum_{i=1}^{N_{\bar \epsilon}} \Pr\{ \mathbb{B}(0,i) \text{ is closed}\} > \bar \epsilon$. This implies that there exists an $\epsilon^{'} \in (0,1)$  and $n_1 \in \{1,2,\ldots,N_{\bar \epsilon}\}$ such that $\Pr\left\{\mathbb{B}(0,n_1) \text{ is}\right.$ $\left. \text{closed}\right\} > \epsilon^{'}$.
Thus, from stationarity of the PPPs, for every $\mathbb{N}\ni r_1 > n_1$ we have,
\begin{eqnarray}a
\epsilon^{'} < \Pr\{\mathbb{B}(0,n_1) \text{ is closed}\} \nonumber
&=&  \Pr\left\{\mathcal{L}_{a,o\not \to n_1} \bigcap \mathcal{V}(r_1)\right\} + \Pr\left\{\mathcal{L}_{a,o\not \to n_1} \bigcap \mathcal{V}^c(r_1)\right\},\nonumber\\
&\leq&\Pr\left\{\mathcal{L}_{a,o\not \to n_1} \bigcap \mathcal{V}(r_1)\right\} + \Pr\left\{\mathcal{V}^c(r_1)\right\} \label{eq:perc_condition_bound}
\end{eqnarray}a
where $\mathcal{L}_{a,o\not \to n_1} := \bigcap_{x \in \mathbb{B}(0,n_1)^c \bigcap \Phi_l} \left\{\mathbf{1}_s\left\{o\rightarrow x, \textbf{S} \right\} = 0\right\}$ and $\mathcal{V}(r_1):=\left\{\exists\right.$ at least one legitimate node in $\left.\mathbb{B}(0,n_1)^c \bigcap \mathbb{B}(0,r_1) \bigcap \Phi_l\right\}$. Also, note that $\Pr\left\{\mathcal{V}^c(r_1)\right\} = \exp\left\{-\lambda_l \pi (r_1^2 - n_1^2)\right\}$. Thus, by choosing $\infty > r_1 > \sqrt{\frac{1}{\lambda_l \pi} \log\frac{2}{\epsilon^{'}} + n_1^2}$, we have $\Pr\left\{\mathcal{V}^c(r_1)\right\} < \epsilon^{'}/2$. Using this, \eqref{eq:perc_condition_bound} can be written as $\epsilon_1 =\epsilon^{'}/2 < \Pr\left\{\mathcal{L}_{a,o\not \to n_1} \bigcap \mathcal{V}(r_1)\right\}$. Since under the condition of the event $\mathcal{V}(r_1)$,  there exists a node $z \in \mathbb{B}(0,n_1)^c \bigcap \Phi_l$ of finite distance $d_1 =\norm{z}_2< r_1  < \infty$, which leads to 
\begin{eqnarray}
\epsilon_{1} < \Pr\left\{\mathcal{L}_{a,o\not \to n_1}\right\} < \Pr\left\{\mathbf{1}_s\left\{o\rightarrow z, \textbf{S} \right\} = 0\right\} = 1 - \Pr\left\{\mathbf{1}_s\left\{o \rightarrow z, \textbf{S}\right\} = 1  \right\}, 
\end{eqnarray}
where we have used the fact that for events $A$ and $B$ such that $A \subseteq B$, $\Pr\{A\} \leq \Pr\{B\}$. Define $n_s = \lceil \frac{d_1}{s} \rceil$. In \figref{fig:percnosplit}, let the outer rectangle and the sequence of squares be defined as $\mathcal{J}_s = [-\sqrt{3} \beta_e s, d_1 + \sqrt{3} \beta_e s] \times [\sqrt{3} \beta_e s, -\sqrt{3} \beta_e s]$ and $\mathcal{K}_s = \bigcup_{i=0}^{n_s-1}S^{q}_i$, where $S^q_i = [is, (i+1)s] \times [-s/2,s/2]$, $i=0,1,2,\ldots,n_s-1$. The dimension of $\mathcal{J}_s$ is $w_s \times (n_s + 2\sqrt{3} \beta_e) s$, where $w_s = (1+ 2\sqrt{3} \beta_e)s$. Further, let $B_s = \{\text{no EDs in the region } \mathcal{J}_s\}$ and $\mathcal{H}_s = \bigcap_{i=0}^{n_s-1}\left\{\text{ $S^q_i$ contains at least}\right.$ $\left.\text{one legitimate node}\right\}$.

If the event $\{\exists \text{ a secure $s$-path shown in Fig. \ref{fig:percnosplit}}\}$ occurs then there is a secure $s$-path. 
Thus, we choose $\omega=s$. Therefore, $\mathbf{1}_s\left\{o \rightarrow z, \textbf{S} \right\} = 1$ if the event $B_s \bigcap \mathcal{H}_s$ occurs, and $\epsilon_{1} < \Pr\left\{\mathbf{1}_s\left\{o\rightarrow z, \textbf{S} \right\} = 0\right\} = 1 - \Pr\left\{\mathbf{1}_s\left\{o \rightarrow z, \textbf{S}\right\} = 1  \right\}$ can be written as,
\begin{eqnarray}
\epsilon_{1} < 1-\Pr\left\{\mathcal{H}_s \cap B_s \right\} \label{eq:epsilon1_inequality}
= 1-e^{-\lambda_e \abs{B_s}}\left(1-e^{-\lambda_l s^2}\right)^{n_s}.
\end{eqnarray}
Recall that $n_s = \lceil\frac{d_1}{s}\rceil$, and the area of the secure region in \figref{fig:percnosplit} is given by $\abs{B_s} = (n_s + 2  \sqrt{3}\beta_e) w_s s$, where $w_s = (1+2\sqrt{3}\beta_e) s$. This implies that for all $s\in (0,\eta/ \sqrt{3}]$, there is percolation if
$\left(1-e^{-\lambda_l s^2}\right)^{n_s}  > (1 - \epsilon_1) \exp{\{ \lambda_e \abs{B_s} \} }$. Equivalently, if the following inequality holds, then the percolation is bound to happen with non-zero probability: if for some $\delta \geq 0$, 
$(1 - \epsilon_1) \exp{\{ \lambda_e \abs{B_s} \} } < e^{-\delta}$, and 
$\lambda_l > \frac{1}{s^2} \left[\log\left( \frac{1}{1-  \exp\left\{\frac{-\delta}{n_s}\right\}}  \right) \right]$.
The first condition implies that $\lambda_e < \frac{1}{\abs{B_s}} \left[\log \left(\frac{1}{1 - \epsilon_1}\right) - \delta\right]$, $0 < \delta < \log \left(\frac{1}{1 - \epsilon_1}\right)$. Taking the ratio of this and the bound on $\lambda_l$, 
and using $s = \eta/\sqrt{3}$, we get
\begin{eqnarray} \label{eq:nosplit_perc_result_app}
\lambda_l > \frac{\left(n_s + 2  \sqrt{3}\beta_e \right) (1+2\sqrt{3}\beta_e) \log\left(\frac{1}{1 - \exp\left\{-  \frac{\delta }{n_s}  \right\}}\right)}{C_{\delta,\epsilon_1}} \lambda_e,
\end{eqnarray}
where $C_{\delta,\epsilon_1} = \left[\log \left(\frac{1}{1 - \epsilon_1}\right) - \delta\right]$ and $n_s = \left \lceil \frac{d_1 \sqrt{3}}{\eta} \right \rceil$. Since we have chosen $s = \frac{\eta}{\sqrt{3}}$, the aforementioned condition implies that with non-zero probability, a giant component with secure $\frac{\eta}{\sqrt{3}}$-path exists, giving the desired result. This completes the proof of \thrmref{thm:percolation}. $\blacksquare$

\subsection{Proof of Corrollary \ref{corr:nopacketsplit}} \label{app:corr_perc_nsp}
The existence of an $\frac{\eta}{\sqrt{3}}$-path implies the existence of a secure path. Thus, from \thrmref{thm:percolation}, we have 
\begin{eqnarray}
\Pr\{\exists \text{ a giant component}\} \geq \Pr\left\{\exists \text{ a giant component consisting of } \frac{\eta}{\sqrt{3}} \text{ paths}\right\} > \epsilon_1,
\end{eqnarray}
from which the result follows. $\Box$

\subsection{Proof of \thrmref{thm:meanhopnosplit}} \label{app:meanhopnosplit}
Let us assume that the conditions of \thrmref{thm:percolation} are satisfied, which implies that with non-zero probability, there exists a giant component consisting of secure $\frac{\eta}{\sqrt{3}}$-paths. Throughout the proof of \thrmref{thm:meanhopnosplit}, a path refers to a secure $\frac{\eta}{\sqrt{3}}$-path unless otherwise stated explicitly. Let 
\begin{eqnarray}
\mathcal{G}_{\text{nsp}} = \left\{\exists \text{ a giant component consisting of } \frac{\eta}{\sqrt{3}}-\text{paths}\right\}.
\end{eqnarray}
Recall that we need to find an upper bound on $\mathbb{E}_{|\mathcal{G}_\text{nsp}} L_\text{nsp}(0,1)$. To accomplish this, consider two nodes $x^*_0 = \arg \min_{x \in \mathcal{G}_\text{nsp}} \norm{x - (0,0)}_2$ and $x^*_1 = \arg \min_{x \in \mathcal{G}_\text{nsp}} \norm{x - (1,0)}_2$. 
It follows from \cite[Lemma 8]{Kong2007} that \emph{almost surely}, $r_0 := \norm{x_0^*}_2< \infty$ and $r_1 :=\norm{x_1^* - (1,0)}_2  < \infty$. This implies that \emph{almost surely} $\norm{x_0^* - x_1^*}_2 < \infty$. Now, we need to find an upper bound on the number of hops along the minimum path in $\mathcal{G}_\text{nsp}$ between $x_0^*$ and $x_1^*$. 
Denote a box of size $b \times b$ centered at $(0,b/2)$ by $\mathcal{B}_b$. Conditioned on the event $\mathcal{G}_\text{nsp}$, there exists a $\delta > 0$ and $b^* > 0$ such that $\Pr_{|\mathcal{G}_\text{nsp}}\{{\rightarrow} \mathcal{B}_{b^*}\} > \delta$ (see \cite{Balister2010}). Recall that the notation ${\rightarrow} \mathcal{B}_{d^*}$ denotes the left to right (secure) crossing of the square $\mathcal{B}_{b^*}$ by an $\frac{\eta}{\sqrt{3}}$-path. Similarly, by symmetry, we have $\Pr_{|\mathcal{G}_\text{nsp}}\{ \leftarrow \mathcal{B}_{b^*}\} = \Pr_{|\mathcal{G}_\text{nsp}}\{{\uparrow}~ \mathcal{B}_{b^*}\} = \Pr_{|\mathcal{G}_\text{nsp}}\{ \downarrow  \mathcal{B}_{b^*}\} > \delta$. For some $\delta_\text{nsp} > 0$, consider a sequence of squares of size $3^j d_{\delta_{\text{nsp}}} \times 3^j d_{\delta_{\text{nsp}}}$ centered at $\frac{\norm{x_0^* - x_1^*}_2}{2}$ denoted by $B_j$, $j=1,2,\ldots$,  where 
$d_{\delta_{\text{nsp}}} : = \max \left\{\norm{x_0^* - x_1^*}_2, \inf\{h>0: \Pr_{|\mathcal{G}_\text{nsp}}\{\rightarrow B_{h}\} > \delta_{\text{nsp}} \right\}.$

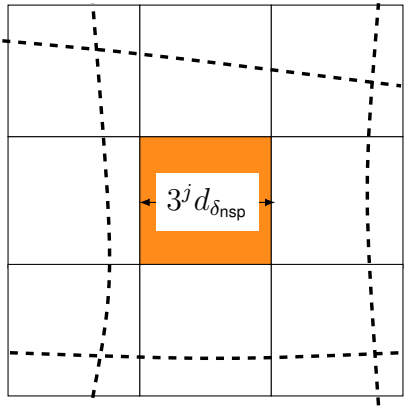
\begin{figure}[ht] 
  \begin{center}
\begin{tikzpicture}
\tikzset{%
    dimen/.style={<->,>=latex,thin,every rectangle node/.style={fill=white,midway,font=\sffamily}},
    symmetry/.style={dashed,thin},
}
\tikzstyle{square_inner}=[fill=white!10!orange,draw=black,minimum height = 1.75cm,minimum width=1.75cm]
\tikzstyle{square_outer}=[fill=white,draw=black,minimum height = 1.75cm,minimum width=1.75cm]
\node[square_inner] at (0,4) (squareinner) {};
\node[square_outer] at (1.75,4) (square1) {};
\node[square_outer] at (-1.75,4) (square2) {};
\node[square_outer] at (0,5.75) (square3) {};
\node[square_outer] at (0,2.3) (square4) {};
\node[square_outer] at (1.75,5.75) (square5) {};
\node[square_outer] at (1.75,2.3) (square6) {};
\node[square_outer] at (-1.75,2.3) (square7) {};
\node[square_outer] at (-1.75,5.75) (square8) {};

\draw[dashed,line width=1.3pt] (-2.6,2) .. controls (.2,1.9) .. (2.6, 2);
\draw[dashed,line width=1.3pt] (2.3,1.3) .. controls (2.1,4.2) .. (2.3, 6.65);
\draw[dashed,line width=1.3pt] (-1.5,1.4) .. controls (-1.2,3) .. (-1.5, 6.65);
\draw[dashed,line width=1.3pt] (-2.7,6.15) .. controls (-1.0,6) .. (2.6, 5.55);
\draw [dimen,<->] ([xshift=0.05cm]squareinner.east) -- ([xshift= 0.0cm]squareinner.west) node  {$3^j d_{\delta_\text{nsp}}$};
\end{tikzpicture}
\caption{A typical square $B_j$ of size $3^j d_{\delta_\text{nsp}} \times 3^j d_{\delta_\text{nsp}}$. 
} \label{fig:delaynoplit}
   \end{center}
\end{figure}

For every $b^*> d_{\delta_{\text{nsp}}}$, $\Pr_{|\mathcal{G}_\text{nsp}}\{ a~\mathcal{B}_{b^*}\} >\delta_{\text{nsp}}$, where $a \in \{{\rightarrow}, {\leftarrow} ,\uparrow,\downarrow \}$. 
Let $E_j$ denote the existence of a secure open circuit consisting of $\frac{\eta}{\sqrt{3}}$-paths inside the annuli $B_j \bigcap B_{j-1}^c$, $j = 1,2,\ldots$. A typical circuit inside the annuli $B_j \bigcap B_{j-1}^c$ is illustrated in \figref{fig:delaynoplit}, where a typical circuit (dashed lines) comprising left (right) to right (left) and lower (upper) to upper (lower) crossings is also shown. 

Suppose if the \emph{minimum delay} path lies entirely inside $B_j$, then none of the events $E_1,E_2,\ldots,E_j$ can occur; otherwise there exists a shorter path leading to a contradiction. Thus, 
\begin{eqnarray}
\Pr_{|\mathcal{G}_\text{nsp}}\left\{\frac{\eta}{\sqrt{3}}\text{-path lies outside } B_j\right\} \leq \Pr_{|\mathcal{G}_\text{nsp}}\{\cap_{i=1}^j E_i^c\}. 
\end{eqnarray}
Consider three consecutive points $a$, $b$ and $c$ on the minimum delay secure path denoted by $L_\text{nsp}(0,1)$. Since the paths in the giant component are secure $\frac{\eta}{\sqrt{3}}$-paths, we have $\eta/\sqrt{3} < d(a,c) \leq 2\eta$. (This is the place where we use the fact that there exists a secure $\eta/\sqrt{3}$-path.) Otherwise, the path $a$ to $c$ exists, resulting in a shorter path compared to $L_\text{nsp}(0,1)$. Thus, we can construct two disjoint circles on $a$ and $c$ of radius $\frac{\eta}{2\sqrt{3}}$. 
In general, along the path $L_\text{nsp}(0,1)$, we can construct at most $L_\text{nsp}/2$ circles of radius $\frac{\eta}{2\sqrt{3}}$ on alternating points, where $L_\text{nsp} = L_\text{nsp}(0,1)$. 
Inside $B_j$, we can have at most 
$\mathbb{A}_j := \left \lceil \frac{12 \times 3^{2j} d_{\delta_{\text{nsp}}}^2}{\pi \eta^2}\right \rceil$
non-overlapping circles of radius $\frac{\eta}{2\sqrt{3}}$. Therefore, if $L_\text{nsp} > 2 \mathbb{A}_j$, none of the events $E_1,E_2,\ldots,E_j$ can occur. Thus,
$\Pr \{L_\text{nsp} > 2 \mathbb{A}_j {|\mathcal{G}_\text{nsp}}\} \leq \Pr\{\cap_{i=1}^j E_i^c{|\mathcal{G}_\text{nsp}}\}
\stackrel{(a)}{\leq} \prod_{i=1}^j \Pr\{E_j^c {|\mathcal{G}_\text{nsp}}\}$, 
where $(a)$ follows from the Fortuin-Kasteleyn-Ginibre (FKG) inequality \cite{Grimmett1999} since $E_1,E_2,\ldots$ are increasing events. We say that an event $A$ is an 
increasing event if $\mathbf{1}_A (G) \leq \mathbf{1}_A (H)$ for every graph $G$ which is a subgraph of $H$.
Conditioned on the existence of a giant component, we have for all $j=1,2,\ldots,$ $\Pr_{|\mathcal{G}_\text{nsp}}\{ a~\mathcal{B}_{d_{\delta_\text{nsp}}}\} >\delta_{\text{nsp}}$, where $a \in \mathcal{C}^\text{(nsp)}_{rs} = \{{\rightarrow}, {\leftarrow}, \uparrow ,\downarrow \}$. Due to this, and using the FKG inequality, we have $\Pr_{|\mathcal{G}_\text{nsp}}\{E_j^c| {\mathcal{G}}\} \leq 1 - \prod_{f \in \mathcal{C}^\text{(nsp)}_{rs}} \Pr_{|\mathcal{G}_\text{nsp}}\{f~\mathcal{B}_{d_{\delta_\text{nsp}}}\}  \leq 1-\delta_{\text{nsp}}^4$. This results in  
$\Pr_{|\mathcal{G}_\text{nsp}}\{L_\text{nsp} > 2 \mathbb{A}_j\} \leq \left(1 - \delta_{\text{nsp}}^4\right)^j$. Therefore, $\mathbb{E}_{|\mathcal{G}_\text{nsp}} L_\text{nsp}$ can be bounded as follows:
\begin{eqnarray}
\mathbb{E}_{|\mathcal{G}_\text{nsp}} L_\text{nsp} = \sum_{i=0}^\infty \Pr\{L_\text{nsp} > i{|\mathcal{G}_\text{nsp}}\}
&\hspace{-8mm}=\hspace{-8mm}& \mathbb{A}_1 + \sum_{j=0}^\infty \sum_{i=\mathbb{A}_j}^{\mathbb{A}_{j+1}} \Pr\{L_\text{nsp} > i{|\mathcal{G}_\text{nsp}}\}
\stackrel{(a)}{\leq}  \mathbb{A}_1 + \sum_{j=0}^\infty \Delta \mathbb{A}_{j+1} p_j\nonumber\\
&\stackrel{(a)}{\leq}& \mathbb{A}_1 +  \sum_{j=0}^\infty \Delta \mathbb{A}_{j+1} \left(1 - \delta_{\text{nsp}}\right)^{j}, \label{eq:mean1}
\end{eqnarray}
where $(a)$ follows from the fact that $\mathbb{A}_{j+1} > \mathbb{A}_j$, $p_j = \Pr\{L_\text{nsp}>\mathbb{A}_j{|\mathcal{G}_\text{nsp}}\}$ and $\Delta \mathbb{A}_{j+1} = \mathbb{A}_{j+1} - \mathbb{A}_{j}$. Also, $\Delta \mathbb{A}_{j+1} \leq \frac{96 d^2_{\delta_{\text{nsp}}}}{\pi \eta^2} 3^{2j} + 1$.

Using this in \eqref{eq:mean1}, we get
\begin{eqnarray}
\mathbb{E}_{|\mathcal{G}_\text{nsp}} L_\text{nsp} \leq \frac{12 d^2_{\delta_{\text{nsp}}}}{\pi \eta^2} + 1 + \frac{96 d^2_{\delta_{\text{nsp}}}}{\pi \eta^2} \sum_{j=0}^\infty 9^j (1- {\delta^4_{\text{nsp}}})^j  + \sum_{j=0}^\infty (1 - {\delta^4_{\text{nsp}}})^j,
\end{eqnarray}
which can further be simplified to get $\mathbb{E}_{|\mathcal{G}_\text{nsp}} L_\text{nsp} \leq\frac{12 d^2_{\delta_{\text{nsp}}}}{\pi \eta^2} \left( \frac{9 {\delta^4_{\text{nsp}}}}{9 {\delta^4_{\text{nsp}}} - 8}\right) + \frac{1 + \delta_{\text{nsp}}^4}{\delta_{\text{nsp}}^4}$. Conditioned on the existence of a giant component, we can choose a ``big enough" square that results in $\delta_{\text{nsp}} > \sqrt[\leftroot{-2}\uproot{2}4]{\frac{8}{9}}$. From this, it follows that $\mathbb{E}_{|\mathcal{G}_\text{nsp}} L_\text{nsp} < \infty$. 
Using the fact that $\mathcal{G}_\text{nsp} \subseteq \mathcal{G}$, we have $\mathbb{E}_{|\mathcal{G}} L_\text{nsp} < \mathbb{E}_{|\mathcal{G}_\text{nsp}} L_\text{nsp}$ which completes the proof of \thrmref{thm:meanhopnosplit}.
$\blacksquare$

\subsection{Proof of Lemma \ref{thm:append_average_delay1}} \label{sec:append_average_delay1}
Let us denote the number of hops corresponding to the minimum delay path by $L_\text{nsp} = L_\text{nsp}(0,1)$ and $T^\text{(nsp)}_1,T^\text{(nsp)}_2,\ldots,T^\text{(nsp)}_{L_\text{nsp}}$ denote the delay for each hop. Thus, the average delay is given by
$\mathbb{E}_{|\mathcal{G}} T_{\text{nsp}} = \mathbb{E}_{|\mathcal{G}} \sum_{i=1}^{L_\text{nsp}} T^\text{(nsp)}_i$,
where the average is with respect to $T^\text{(nsp)}_1,T^\text{(nsp)}_2,\ldots,T^\text{(nsp)}_{L_\text{nsp}}$ and $L_\text{nsp}$. We know that any hop in the giant component is no larger than ${\eta}$. Thus, using \eqref{eq:averagedelayzdistance}, we have \\ $\mathbb{E}_{|\mathcal{G}}\{T^\text{(nsp)}_{i}\} \leq  \exp\left({\eta^2}\left(\lambda_e \pi \beta_e^2  + \lambda_l \frac{p}{1-p} \pi \gamma(\beta_l)\right)\right)$ for all $i=1,2,\ldots,L_\text{nsp}$. Note that, for ``sufficiently" large $\eta$, with high probability there exists a single link between the source and the destination of unit distance in which case $\eta$ can be replaced by a constant. Also, conditioned on the giant component, there are no EDs around the path. Thus, the bound can be tightened by using $\lambda_e = 0$. We ignore this since we are interested only in the scaling behavior. Using this, we have the following bound for the average delay:
\begin{eqnarray}
\mathbb{E}_{|\mathcal{G}} T_{\text{nsp}} = \mathbb{E}_{L_\text{nsp}{|\mathcal{G}}} \sum_{i=1}^{L_\text{nsp}} \mathbb{E}\{T_i\},
\leq \mathbb{E}_{|\mathcal{G}} L_\text{nsp} \times \exp\left({\eta^2}\left(\lambda_e \pi \beta_e^2  + \lambda_l \frac{p}{1-p} \pi \gamma(\beta_l)\right)\right).
\end{eqnarray}
Finally, using the upper bound on $\mathbb{E}_{|\mathcal{G}} L_\text{nsp}$ from \thrmref{thm:meanhopsplit}, we get the desired result. $\blacksquare$

\subsection{Proof of \thrmref{thm:percolation_splitting}} \label{app:percolation_splitting}
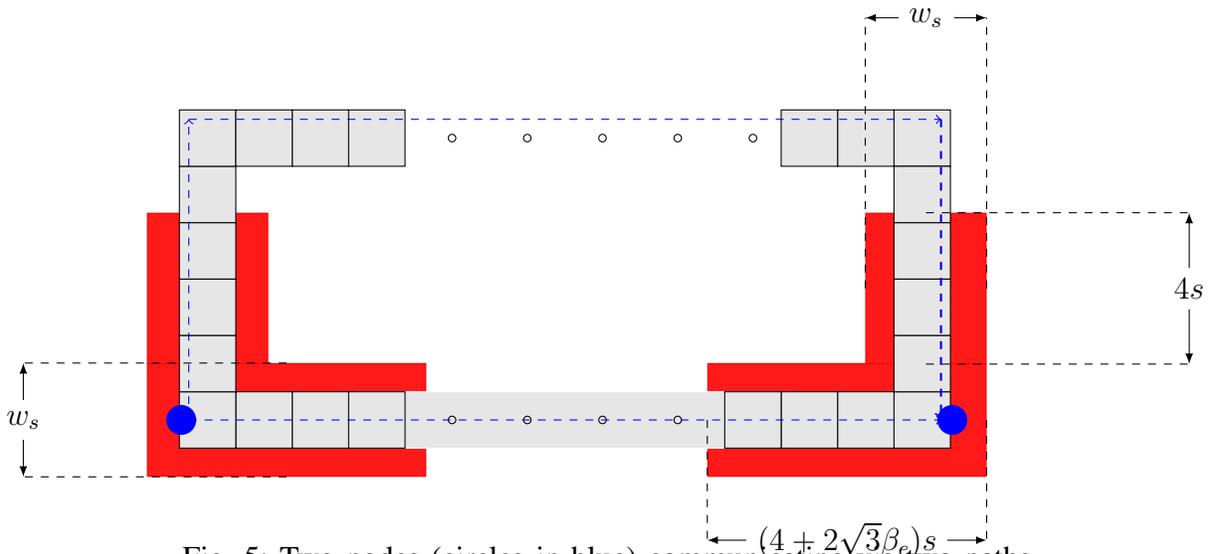
\begin{figure}[ht]
  \begin{center}
\begin{tikzpicture}[scale=1]
\tikzset{%
    dimen/.style={<->,>=latex,thin,every rectangle node/.style={fill=white,midway,font=\sffamily}},
    symmetry/.style={dashed,thin},
}

\tikzstyle{lightGrayBoxStyle}=[fill=gray!20,draw=black,minimum height = 0.75cm,minimum width=0.75cm]
\tikzstyle{outerbox1}=[fill= white!10!red,draw=red,minimum height=1.5cm,minimum width=3.70cm]
\tikzstyle{outerbox2}=[fill= white!10!red,draw=red,minimum height=2.0cm,minimum width=1.6cm]
\tikzstyle{lightGrayBoxthin}=[fill=gray!20,draw=white,minimum height = 0.75cm,minimum width=5.80cm]

\node[outerbox1] at (1.8,0.75) (oh1) {};
\node[outerbox2] at (0.75,2.50) (ov1) {};
\node[outerbox1] at (9.25,0.75) (oh2) {};
\node[outerbox2] at (10.30,2.50) (ov2) {};
\node[lightGrayBoxthin] at (5.6,0.75) (horizontaltile) {};

\node[lightGrayBoxStyle] at (10.25,0.75) (b9) {};
\node[lightGrayBoxStyle] at (9.5,0.75) (b8) {};
\node[lightGrayBoxStyle] at (8.75,0.75) (b7) {};
\node[lightGrayBoxStyle] at (8.0,0.75) (b6) {};
\node[lightGrayBoxStyle] at (3.0,0.75) (b4) {};
\node[lightGrayBoxStyle] at (2.25,0.75) (b3) {};
\node[lightGrayBoxStyle] at (1.5,0.75) (b2) {};
\node[lightGrayBoxStyle] at (0.75,0.75) (b1) {};

\node[lightGrayBoxStyle] at (0.75,1.5) (bv4) {};
\node[lightGrayBoxStyle] at (0.75,2.25) (bv3) {};
\node[lightGrayBoxStyle] at (0.75,3.0) (bv2) {};
\node[lightGrayBoxStyle] at (0.75,3.75) (bv1) {};
\node[lightGrayBoxStyle] at (0.75,4.5) (bv0) {};

\node[lightGrayBoxStyle] at (10.25,4.5) (bh6) {};
\node[lightGrayBoxStyle] at (9.5,4.5) (bh5) {};
\node[lightGrayBoxStyle] at (8.75,4.5) (bh4) {};

\node[lightGrayBoxStyle] at (3.0,4.5) (bh3) {};
\node[lightGrayBoxStyle] at (2.25,4.5) (bh2) {};
\node[lightGrayBoxStyle] at (1.5,4.5) (bh1) {};

\node[lightGrayBoxStyle] at (10.25,1.5) (bvr4) {};
\node[lightGrayBoxStyle] at (10.25,2.25) (bvr3) {};
\node[lightGrayBoxStyle] at (10.25,3.0) (bvr2) {};
\node[lightGrayBoxStyle] at (10.25,3.75) (bvr1) {};

\draw (4,4.5) circle (0.05cm);
\draw (5,4.5) circle (0.05cm);
\draw (6,4.5) circle (0.05cm);
\draw (7,4.5) circle (0.05cm);
\draw (8,4.5) circle (0.05cm);

\draw (4,0.75) circle (0.05cm);
\draw (5,0.75) circle (0.05cm);
\draw (6,0.75) circle (0.05cm);
\draw (7,0.75) circle (0.05cm);
\draw [dimen,<->] ([xshift=-3.5cm]oh1.south) -- ([xshift=-3.5cm]oh1.north) node  {$w_s$};
\draw [dimen,<->] ([xshift=3.5cm]ov2.north) -- ([xshift=3.5cm]ov2.south) node  {$4s$};

\draw[dashed] ([xshift=-0.0cm]oh1.south) --([xshift=-3.5cm]oh1.south);
\draw[dashed] ([xshift=-0.0cm]oh1.north) --([xshift=-3.5cm]oh1.north);
\draw[dashed] ([xshift=-0.0cm]ov2.north) --([xshift=3.5cm]ov2.north);
\draw[dashed] ([xshift=-0.0cm]ov2.south) --([xshift=3.5cm]ov2.south);

\draw (0.4,0.75) node[circle,fill=blue]{} circle (0.01cm);
\draw (10.65,0.75) node[circle,fill=blue]{} circle (0.01cm);
\draw[blue,dashed,->] (0.5,0.75) -- (10.5,0.75);
\draw[blue,dashed,->] (0.5,0.75) -- (0.5,4.75);
\draw[blue,dashed,->] (0.5,4.75) -- (10.5,4.75);
\draw[blue,dashed,thick,->] (10.5,4.75) -- (10.5,0.75);
\draw [dimen,<->] ([yshift=3.6cm]ov2.west) -- ([yshift=3.6cm] ov2.east) node {$w_s$};
\draw[dashed] ([yshift=0.0cm]ov2.west) --([yshift=3.6cm]ov2.west);
\draw[dashed] ([yshift=0.0cm]ov2.east) --([yshift=3.6cm]ov2.east);
\draw [dimen,<->] ([yshift=-1.6cm]oh2.west) -- ([yshift=-1.6cm] oh2.east) node {$(4+2\sqrt{3}\beta_e)s$};
\draw[dashed] ([yshift=-0.0cm]oh2.west) --([yshift=-1.6cm]oh2.west);
\draw[dashed] ([yshift=-0.0cm]oh2.east) --([yshift=-1.6cm]oh2.east);
\end{tikzpicture}
\vspace{-0.6cm}
\caption{Two nodes (circles in blue) communicating via two paths.}
\label{fig:percspit}
  \end{center}
\end{figure}

Similar to the proof of Theorem \ref{thm:percolation}, we prove this theorem by contradiction. Suppose there is no percolation, {\ie},
$\Pr\{\exists ~ C \subseteq \Phi_l : \abs{C} = \infty~\text{and}~\forall(x,y)\subseteq \Phi_l,~~\mathbf{1}_{s}(x {\rightrightarrows} y,\mathbf{S})=1\} = 0$,
where for some $s>0$ (to be chosen later), the event $\{\mathbf{1}_{s}(x {\rightrightarrows} y,\mathbf{S})=1\}$ indicates that there exists a two-secure $s$-path, as explained in \secref{sec:packetsplit}. Note that, the entire path between nodes need not be secure. Only secure regions around the source and the destination nodes suffice, as illustrated in \figref{fig:percspit}, where the union of secure regions marked in red (along with the corresponding gray region) to the left and right is denoted by $B_s$. The packets can be forwarded securely if the region $B_s$ does not contain any EDs. Along the path, each tile is of size $s\times s$. The outer box is of size $w_s = (1+ 2\sqrt{3} \beta_e)s$.

By an argument similar to the proof of Theorem \ref{thm:percolation}, 
we can show that there exists an $\epsilon_{2}\in (0,1)$, and a node $z \in \mathbb{B}(0,d_2)^c \bigcap \Phi_l$ at a distance $d_2 = \norm{z}_2 < \infty$ such that
\begin{eqnarray}
\epsilon_{2} &<& \Pr\left\{\nexists \text{ a two-secure $s$-path between }{o} \text{ and any node in } \mathbb{B}(0,d_2)^c \bigcap \Phi_l\right\}  \nonumber  \\
&\leq& \Pr\{\mathcal{O}_1^c \bigcap \mathcal{O}_2^c\}
\leq \min\{\Pr\{\mathcal{O}_1^c\}, \Pr\{\mathcal{O}_2^c\}\},\label{eq:epsilon2bound}
\end{eqnarray}
where 
\begin{eqnarray}
\mathcal{O}_1 &=& \left\{\exists \text{ a secure}~s-\text{path between}~o~\text{and the node}~z~\text{in}~\mathbb{B}(0,d_2)^c \bigcap \Phi_l\right\}, \\
\mathcal{O}_2 &=& \left\{\exists \text{ a split-secure}~s-\text{path between}~o~\text{and the node}~z~\text{in}~\mathbb{B}(0,d_2) c \bigcap \Phi_l\right\}.
\end{eqnarray}
Here, split $s$-path refers to a two-secure path which is also a $s$-path.
Now, \eqref{eq:epsilon2bound} implies that
\begin{eqnarray} \label{eq:o1o2events}
\epsilon_2 < \Pr\{\mathcal{O}_1^c\} \text{  and  } \epsilon_2 < \Pr\{\mathcal{O}_2^c\},
\end{eqnarray}
where, for $s = \frac{\eta}{\sqrt{3}}$, the first inequality coincides with \eqref{eq:epsilon1_inequality} if $\epsilon_1$ is replaced by $\epsilon_2$, and $d_1$ is replaced by $d_2$. Thus, using \eqref{eq:nosplit_perc_result_app},
we have
\begin{eqnarray} \label{eq:lambdal_cond_spilt1}
\lambda_l > \frac{\left(\left \lceil \frac{d_2 \sqrt{3}}{\eta} \right \rceil + 2  \sqrt{3}\beta_e \right) (1+2\sqrt{3}\beta_e) \log\left(\frac{1}{1 - \exp\left\{-\delta \left \lceil\frac{d_2 \sqrt{3}}{\eta} \right\rceil\right\}}\right)}{C_{\delta,\epsilon_2}} \lambda_e, 
\end{eqnarray}
where $C_{\delta,\epsilon_2} = \left[\log \left(\frac{1}{1 - \epsilon_2}\right) - \delta\right]$. (With a slight abuse of notation, we have used the same notation $\delta$ as in \eqref{eq:epsilon1_inequality}.)

From Fig. \ref{fig:percspit}, it is clear that $\mathcal{O}_2$ occurs if the regions marked in red denoted \\ $B_s:=\left(o,((4+2\sqrt{3}\beta_e)s,w_s)\right) \bigcup \left(o,(w_s,(5+2\sqrt{3}\beta_e)s)\right)$ and $B_d:=\{-(x + d_1 + 2\sqrt{3}\beta_es): x \in B_s\}$ contain no EDs, 
and each square of size $s \times s$ along the two paths shown contains at least one legitimate node. Thus, the second expression in \eqref{eq:o1o2events} can be written as follows:
\begin{eqnarray}
\epsilon_{2} &<& 1-\Pr\left\{ \left\{o {\stackrel{(\textbf{R})}{\rightrightarrows}} (d_2,0)\right\} \cap \{B_s \text{ and } B_d \text{ are secure}\}\right\}
= 1-e^{-2 \lambda_e \abs{B_s}}\mathcal{J}^{2(n_s+4)}, \label{eq:epsilon2bound2}
\end{eqnarray}
where $\mathcal{J} = \left(1-e^{-\lambda_l s^2}\right)$, $n_s = \lceil\frac{d_2}{s}\rceil$ for some $d_2 > n_0$, and the area of the secure region in the figure is given by 
$\abs{B_s} = 2s^2(1 + 2\sqrt{3}\beta_e) (4 + \sqrt{3}\beta_e)$. The notation $\left\{o {\stackrel{(\textbf{R})}{\rightrightarrows}} (d_2,0)\right\}$ denotes that there exists a rectangluar path in the region $\mathcal{R}_{1} \bigcap \mathcal{R}_{2}^c$, where 
\begin{eqnarray}
\mathcal{R}_{1} &=& \{(\sqrt{3}\beta_e s, \sqrt{3}\beta_e s) \times (\sqrt{3}\beta_e s + d_2, (\sqrt{3}\beta_e + 6) s)\}, \\
\mathcal{R}_{2} &=& \{\left((\sqrt{3}\beta_e + 1) s,(\sqrt{3}\beta_e + 1) s\right) \times \left((\sqrt{3}\beta_e + 1) s + d_2,4s\right), 
\end{eqnarray}
as shown in \figref{fig:percspit}. The areas of the two regions $B_s$ and $B_d$ are same. In \eqref{eq:epsilon2bound2}, the exponent ${2(n_s+4)}$ corresponds to 
the number of square regions of size $s \times s$ across the two paths shown in \figref{fig:percspit}. Thus, there is percolation if 
$(1 - \epsilon_2) \exp\{2 \lambda_e \abs{B_s}\} >  \left(1-e^{-\lambda_l s^2}\right)^{2(n_s+4)}$
provided there exists a $\lambda_l$ such that the right hand side above is less than one, {\ie}, for every $\delta \geq 0$, 
\begin{eqnarray}
\left(1-e^{-\lambda_l s^2}\right)^{2(n_s+4)} < \exp\{-\delta\} \implies \lambda_l > \frac{1}{s^2} \left[\log\left(\frac{1}{1-\exp\left\{- \frac{\delta}{2\left(n_s + 4 \right)}\right\}}\right)\right], 
\end{eqnarray}
and the left hand side is less than $\delta$, {\ie},  for $0 \leq \delta < \log \left(\frac{1}{1 - \epsilon_2}\right) $, $\lambda_e < \frac{1}{2 \abs{B_s}} \left[\log \left(\frac{1}{1 - \epsilon_2}\right) - \delta\right]$. Taking the ratio of these two inequalities and using $s = \eta/\sqrt{3}$ results in 
\begin{eqnarray}
\lambda_l > \frac{4 \left(4 +  \sqrt{3} \beta_e\right) (1 + 2 \sqrt{3} \beta_e) }{C_{\delta, \epsilon_2}} \log\left(\frac{1}{1 - \exp\left\{-\frac{\delta}{2 \left(n_s + 4 \right)}\right\} } \right) \lambda_e,
\end{eqnarray}
where $n_s =\left\lceil\frac{\sqrt{3} d_2}{\eta} \right \rceil $ and $C_{\delta,\epsilon_2} = \left[\log \left(\frac{1}{1 - \epsilon_2}\right) - \delta\right]$ for any $0 \leq \delta < \log \left(\frac{1}{1 - \epsilon_2}\right)$. Combining this with \eqref{eq:lambdal_cond_spilt1}, and the fact that the existence of an $\frac{\eta}{\sqrt{3}}-$path implies the existence of a two-secure path, proves \thrmref{thm:percolation_splitting}. $\blacksquare$

\subsection{Proof of Theorem \ref{thm:meanhopsplit} }\label{app:meanhopsplit}
Let us assume that the conditions of Theorem \ref{thm:percolation_splitting} are satisfied, which implies that with non-zero probability, there exists a giant component consisting of two-secure $\frac{\eta}{\sqrt{3}}$-paths. Recall that a two-secure path consists of a combination of secure direct paths and secure split-paths. Let 
\begin{eqnarray}
\mathcal{G}_{\text{sp}} = \left\{\exists~\text{a giant component with two-secure}~\frac{\eta}{\sqrt{3}}-\text{paths} \right\}.
\end{eqnarray}
Consider two nodes $x^*_0 := \arg \min_{x \in \mathcal{G}_\text{sp}} \norm{x - (0,0)}_2$ and $x^*_1 := \arg \min_{x \in \mathcal{G}_\text{sp}} \norm{x - (1,0)}_2$. It follows from \cite[Lemma 8]{Kong2007} that $r_0 := \norm{x_0^* - (0,0)}_2< \infty$ and $r_1 :=\norm{x_1^* - (1,0)}_2  < \infty$ \emph{almost surely}. This implies that $\norm{x_0^* - x_1^*}_2 < \infty$ \emph{almost surely}. Now, we need to find an upper bound on the number of hops along the minimum path in $\mathcal{G}_\text{sp}$ between $x_0^*$ and $x_1^*$. 
Denote a box of size $b \times b$ centered at $(0,b/2)$ by $\mathcal{B}_b$. 
Conditioned on the event $\mathcal{G}_{\text{sp}}$, there exists a $\delta > 0$ and $d^* > 0$ such that $\Pr_{|\mathcal{G}_{\text{sp}}}\{ {\rightrightarrows} \mathcal{B}_{d^*}\} > \delta$. For convenience, we use ${\rightrightarrows} \mathcal{B}_{d^*}$ to denote the left to right crossing of the square $\mathcal{B}_{d^*}$, where the paths are two-secure $\frac{\eta}{\sqrt{3}}$-paths. By symmetry, we have 
\begin{eqnarray}
\Pr_{|\mathcal{G}_{\text{sp}}}\{ {\leftleftarrows}~ \mathcal{B}_{d^*}\} = \Pr_{|\mathcal{G}_{\text{sp}}}\{ \upuparrows \mathcal{B}_{d^*}\} = \Pr_{|\mathcal{G}_{\text{sp}}}\{ {\downdownarrows}~ \mathcal{B}_{d^*}\} > \delta.
\end{eqnarray}

\begin{figure}[ht] 
 \begin{center}
\begin{tikzpicture}[scale=.4]
\tikzset{%
    dimen/.style={<->,>=latex,thin,every rectangle node/.style={fill=white,midway,font=\sffamily}},
    symmetry/.style={dashed,thin},
}
\tikzstyle{rectangle}=[fill=white!20,draw=black,minimum height=3.5cm,minimum width=10cm]
\shade[shading=ball, ball color=blue]  (-5,0) circle (.3);
\shade[shading=ball, ball color=blue]  (4,0) circle (.3);
\shade[shading=ball, ball color=blue]  (9,0) circle (.3);
\shade[shading=ball, ball color=blue]  (15,0) circle (.3);
\shade[shading=ball, ball color=blue] (1.6,-4.4) rectangle (2.4,-5);
\node (y) at (1.6,-5.5) {Node $1$};
\shade[shading=ball, ball color=blue] (7,-4.4) rectangle (7.8,-5);
\node (y) at (7.8,-5.5) {Node $2$};
\shade[shading=ball, ball color=blue]  (-2,-2) circle (.3);
\shade[shading=ball, ball color=blue]  (-2, 2) circle (.3);

\shade[shading=ball, ball color=blue]  (13,-3) circle (.3);
\shade[shading=ball, ball color=blue]  (13, 2) circle (.3);

\draw[line width=1pt] (-5, 0) .. controls(-2,1) .. (4, 0);
\draw[line width=1pt] (4, 0) .. controls(5,1) .. (9, 0);
\draw[line width=1pt] (9, 0) .. controls(13,1) .. (15, 0);
\draw[dashed] (9, 0) .. controls(13,-2) .. (15, 0);

\draw[line width=1pt] (-2, -2) .. controls(1,1) .. (-2, 2);
\draw[dashed] (-2, -2) .. controls(-3,1) .. (-2, 2);
\node (z) at (-2.4, -2.5) {$L_1$};
\node (x) at (-2.5, 2.5) {$L_2$};

\draw[line width=1pt] (13, -3) .. controls(14,1) .. (13, 2);
\draw[dashed] (13, -3) .. controls(12,1) .. (13, 2);
\node (y) at (13.6, -3.5) {$R_1$};
\node (w) at (13.6, 2.4) {$R_2$};

\draw[line width=1pt] (13, -3) .. controls(11,-4) .. (7.8,-4.7);
\draw[line width=1pt] (13, 2) .. controls(13,3) .. (13, 4);
\draw[line width=1pt] (-2, -2) .. controls(-2,-3) .. (1.9,-4.8);
\draw[line width=1pt] (-2, 2) .. controls(-2,3) .. (-2, 4);

\draw[dashed] (-2, 4) .. controls(7,4) .. (13, 4);


\node (a) at (-3.2, 1) {A};
\node (b) at (0.6, 0.9) {B};

\node (c) at (11.8, 1) {C};
\node (d) at (14,1) {D};

\node (e) at (12.3, -1.7) {E};
\node (f) at (13.7,-1.7) {F};
\draw[-] (-7, -5) .. controls(-7,3) .. (-7, 3);
\draw[-] (-7, 3) .. controls(3,3) .. (16, 3);
\draw[-] (16, 3) .. controls(16,2) .. (16, -5);
\node (g) at (4.3, 2.2) {Box $B_j$};
\end{tikzpicture} 
\caption{A typical path between two nodes denoted Node $1$ and Node $2$ inside the box $B_j$.} 
\label{fig:cross_over_points}
\end{center} 
\end{figure}
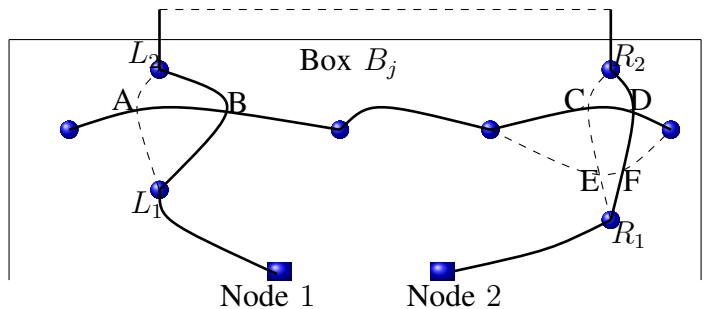

For some $\delta_{\text{sp}} >0$, consider a sequence of squares of size $3^j d_{\delta_{\text{nsp}}} \times 3^j d_{\delta_{\text{nsp}}}$ centered at $\frac{\norm{x_0^* - x_1^*}_2}{2}$ denoted by $B_j$, $j=1,2,\ldots$,  where 
$d_{\delta_{\text{sp}}} : = \max \left\{d, \inf\{\bar d: \Pr\{\rightrightarrows B_{\bar d}\} > \delta_{\text{sp}}\} \right\}.$
Note that, for every $d^*> d_{\delta_{\text{sp}}}$, $\Pr_{|\mathcal{G}_{\text{sp}}}\{ a~\mathcal{B}_{d^*}\} >\delta_{\text{sp}}$, where $a \in \{{\rightrightarrows}, \leftleftarrows,\upuparrows,\downdownarrows\}$. 
Let $E_j$ denote the existence of a secure open rectangle circuit inside the annuli $B_j \bigcap B_{j-1}^c$, $j = 1,2,\ldots$. 
Suppose the minimum delay path lies entirely inside $B_j$. Then none of the events $E_1,E_2,\ldots,E_j$ can occur; otherwise there exists an even shorter path, leading to a contradiction. See \figref{fig:cross_over_points}, where one of the path goes out of the box $B_j$ and the other takes a path along the circuit, {\ie}, Node $1$ to $L_1$ to $(A,B)$ to $(C,D)$ to $(E,F)$ to $R_1$ to Node $2$. The path along the circuit is shorter compared to the one that goes out of the box. The crossover to the left and right correspond to a secure split-path intersecting with a secure direct-path and split-paths crossing each other, respectively.


From \figref{fig:cross_over_points} it is easy to see that the shortest path cannot go out of the circuit. 
Thus,
\begin{eqnarray}
\Pr_{|\mathcal{G}_{\text{sp}}}\left\{\text{two-secure } \frac{\eta}{\sqrt{3}}\text{-path lies outside of } B_j\right\} \leq \Pr_{|\mathcal{G}_{\text{sp}}}\{\cap_{i=1}^j E_i^c\}.
\end{eqnarray} 
Consider three consecutive points $a$, $b$ and $c$ on $L_\text{sp}(0,1)$. Next, for both split and direct paths, we know from Theorem \ref{thm:percolation_splitting} that $2 \eta > d(a,c) > \eta/\sqrt{3}$. 
Thus, we can construct two disjoint circles on $a$ and $c$ of radius $\frac{\eta}{2 \sqrt{3}}$ or more. 
In general, along the path $L_\text{sp} := L_\text{sp}(0,1)$, we can construct at most $L_\text{sp}/2$ disjoint circles of radius $\frac{\eta}{2\sqrt{3}}$ on alternating points. Inside $B_j$, we can have at most 
$\mathbb{A}_j := \left \lceil \frac{12 \times 3^{2j} d_{\delta_{\text{sp}}}^2}{\pi \eta^2}\right \rceil$
non-overlapping circles of radius $\frac{\eta}{2\sqrt{3}}$. Therefore, if $L_\text{sp} > 2 \mathbb{A}_j$, none of the events $E_1,E_2,\ldots,E_j$ can occur. Thus,
$\Pr_{|\mathcal{G}_{\text{sp}}}\{L_\text{sp} > 2 \mathbb{A}_j\} \leq \Pr_{|\mathcal{G}_{\text{sp}}}\{\cap_{i=1}^j E_i^c\}
\stackrel{(a)}{\leq} \prod_{i=1}^j \Pr_{|\mathcal{G}_{\text{sp}}}\{E_j^c\}$ 
where $(a)$ follows from the FKG inequality, since $E_1,E_2,\ldots$ are increasing events.
Conditioned on the existence of a giant component, we have for all $j=1,2,\ldots,$ $\Pr\{ a~\mathcal{B}_{d^*}\} >\delta_{\text{nsp}}$, where $a \in \mathcal{C}_{rs}^\text{sp} :=\{{\rightrightarrows}, {\leftleftarrows}, \upuparrows,\downdownarrows \}$. Due to this, and using the FKG inequality, we have $\Pr_{|\mathcal{G}_{\text{sp}}}\{E_j^c\} \leq 1 - \prod_{f \in \mathcal{C}_{rs}^\text{sp}} \Pr_{|\mathcal{G}_{\text{sp}}}\{f~\mathcal{B}_{d^*}\}  \leq 1-\delta_{\text{nsp}}^4$. This results in
$\Pr_{|\mathcal{G}_{\text{sp}}}\{L_\text{sp} > 2 \mathbb{A}_j\} \leq \left(1 - \delta_{\text{sp}}^4\right)^j$ 
which leads to
\begin{eqnarray}
\mathbb{E}_{|\mathcal{G}_{\text{sp}}} L_\text{sp} := \sum_{i=0}^\infty \Pr_{|\mathcal{G}_{\text{sp}}}\{L_\text{sp} > i\}
\leq \mathbb{A}_1 +  \sum_{j=0}^\infty (\mathbb{A}_{j+1} - \mathbb{A}_{j}) \left(1 - \delta_{\text{sp}}\right)^{j}, 
\end{eqnarray}
which is obtained in a manner similar to \eqref{eq:mean1} in the proof of Theorem \ref{thm:meanhopnosplit}. Using
$\mathbb{A}_{j+1} - \mathbb{A}_{j} \leq \frac{96 d^2_{\delta_{\text{sp}}}}{\pi \eta^2} 3^{2j} + 1$, we can write 
\begin{eqnarray}
\mathbb{E}_{|\mathcal{G}_{\text{sp}}} L_\text{sp} \leq \frac{12 d^2_{\delta_{\text{sp}}}}{\pi \eta^2} + 1 + \frac{96 d^2_{\delta_{\text{sp}}}}{\pi \eta^2} \sum_{j=0}^\infty 9^j (1- {\delta^4_{\text{sp}}})^j  + \sum_{j=0}^\infty (1 - {\delta^4_{\text{sp}}})^j 
= \frac{12 d^2_{\delta_{\text{sp}}}}{\pi \eta^2} \left( \frac{9 {\delta^4_{\text{sp}}}}{9 {\delta^4_{\text{sp}}} - 8}\right) + \frac{1 + \delta_{\text{sp}}^4}{\delta_{\text{sp}}^4}.
\end{eqnarray}
Conditioned on $\mathcal{G}_\text{sp}$, we can choose a ``sufficiently large" square such that $\delta_{\text{sp}} > \sqrt[\leftroot{-2}\uproot{2}4]{\frac{8}{9}}$. Using this, it is easy to see that $\mathbb{E}_{|\mathcal{G}_{\text{sp}}} L_\text{sp} < \infty$. 
The fact that $\mathcal{G}_\text{sp} \subseteq \mathcal{G}$ completes the proof of \thrmref{thm:meanhopsplit}. $\blacksquare$

\bibliographystyle{IEEEtran}
\bibliography{IEEEabrv,TCOM2019}

\end{document}